\documentclass[12pt,preprint]{aastex}
\usepackage{color}

\newcommand{\degree}{{^\circ}}

\newcommand{\cmnosp}{{\rm cm}}
\newcommand{\kmnosp}{{\rm km}}
\newcommand{\aunosp}{{\rm AU}}
\newcommand{\pcnosp}{{\rm pc}}

\newcommand{\gramnosp}{{\rm g}}

\newcommand{\dynnosp}{{\rm dyn}}

\newcommand{\msunnosp}{{\rm M_\odot}}

\newcommand{\secondnosp}{{\rm s}}
\newcommand{\gaussnosp}{{\rm G}}
\newcommand{\muGnosp}{{\mu\rm G}}

\newcommand{\cm}{{\rm\,\cmnosp}}

\newcommand{\au}{{\rm\,\aunosp}}
\newcommand{\pc}{{\rm\,\pcnosp}}

\newcommand{\dyn}{{\rm\,\dynnosp}}

\newcommand{\msun}{{\rm\,\msunnosp}}

\newcommand{\second}{{\rm\,\secondnosp}}

\newcommand{\muG}{{\rm\,\muGnosp}}

\newcommand{\kmsnosp}{{\kmnosp\second^{-1}}}
\newcommand{\cmsnosp}{{\cmnosp\second^{-1}}}
\newcommand{\etaunitnosp}{{\cmnosp^2\second^{-1}}}
\newcommand{\rhounitnosp}{{\gramnosp\cm^{-3}}}

\newcommand{\kms}{{\rm\,\kmsnosp}}
\newcommand{\cms}{{\rm\,\cmsnosp}}
\newcommand{\etaunit}{{\rm\,\etaunitnosp}}
\newcommand{\rhounit}{{\rm\,\rhounitnosp}}

\newcommand{\ct}{\citealt}

\begin{document}

\title{On the Role of Pseudodisk Warping and Reconnection in Protostellar
Disk Formation in Turbulent Magnetized Cores}

\author{Zhi-Yun Li\altaffilmark{1}, Ruben
Krasnopolsky\altaffilmark{2,3}, Hsien Shang\altaffilmark{2,3}, Bo
Zhao\altaffilmark{1}}
\altaffiltext{1}{University of Virginia, Astronomy Department, Charlottesville, USA}
\altaffiltext{2}{Academia Sinica, Institute of Astronomy and Astrophysics, Taipei, Taiwan}
\altaffiltext{3}{Academia Sinica, Theoretical Institute for Advanced Research in Astrophysics, Taipei, Taiwan}

\shorttitle{\sc Protostellar Disk Formation}

\begin{abstract}

The formation of rotationally supported protostellar disks is
suppressed in ideal MHD in non-turbulent cores with aligned magnetic
field and rotation axis. A promising way to resolve this so-called
``magnetic braking catastrophe'' is through turbulence. The
reason for the turbulence-enabled disk formation is usually attributed
to the turbulence-induced magnetic reconnection, which is thought to
reduce the magnetic flux accumulated in the disk-forming region. We
advance an alternative interpretation, based on magnetic
decoupling-triggered reconnection of severely pinched field lines close to the central protostar
and turbulence-induced warping of the pseudodisk of Galli
and Shu. Such reconnection weakens the central split magnetic monopole
that lies at the heart of the magnetic braking catastrophe under flux
freezing. We show,
through idealized numerical experiments,
that the pseudodisk can be strongly warped, but not
completely destroyed, by a subsonic or sonic turbulence. The warping
decreases the rates of angular momentum removal from the pseudodisk by
both magnetic torque and outflow, making it easier to form a
rotationally supported disk. More importantly, the warping of the
pseudodisk out of the disk-forming, equatorial plane greatly
reduces the amount of magnetic flux threading the circumstellar,
disk-forming region, further promoting disk formation.
The beneficial effects of pseudodisk warping can also be
achieved by a misalignment between the magnetic field and
rotation axis. These two mechanisms of disk formation, enabled by
turbulence and field-rotation misalignment respectively, are thus
unified. We find that the disks formed in turbulent magnetized cores
are rather thick and significantly magnetized. Implications of these
findings, particularly for the thick young disk inferred in L1527, are
briefly discussed.
\end{abstract}


\section{Introduction}

How disks form is a longstanding, unsolved problem in star formation
(\ct{Bodenheimer1995}).
Observationally, it has been difficult to determine when and how the
disks first come into existence during the star formation process.
Although disks are routinely observed around evolved Class II young
stellar objects (see \ct{WilliamsCieza2011} for a review) and
increasingly around younger Class I sources (e.g., \ct{Brinch+2007};
\ct{Jorgensen+2009}; \ct{Lee2011}; \ct{Takakuwa+2012};
\ct{Harsono+2014}; \ct{Lindberg+2014}), observations of the youngest disks have been
hampered by the emission from their massive envelope. Nevertheless,
rotationally supported disks are beginning to be detected around
deeply embedded, Class 0 protostars (\ct{Tobin+2012}; \ct{Murillo+2013};
N. Ohashi, priv.\ comm.). This impressive observational progress
is expected to accelerate in the near future, as ALMA becomes fully
operational.

Theoretically, disk formation is complicated by magnetic fields,
which are observed in dense, star-forming, cores of molecular
clouds (see \ct{Crutcher2012} for a review). In the simplest case
of a non-turbulent core with the magnetic field aligned with the
rotation axis, both analytic considerations and numerical simulations
have shown that the formation of a rotationally supported disk
(RSD hereafter) is suppressed, in the ideal MHD
limit, by a realistic magnetic field (corresponding to a
dimensionless mass-to-flux ratio of $\lambda
\sim$ a few; \ct{TrolandCrutcher2008}) during the protostellar mass
accretion phase through magnetic braking
(\ct{Allen+2003}; \ct{Galli+2006}; \ct{PriceBate2007};
\ct{MellonLi2008}; \ct{HennebelleFromang2008}; \ct{DappBasu2010};
\ct{Seifried+2011}; \ct{Santos-Lima+2012}). This suppression of RSDs
is termed the ``magnetic braking catastrophe.''

There are a number of mechanisms proposed in the literature to
overcome the catastrophic braking, including (1) non-ideal
MHD effects (ambipolar diffusion, the Hall effect and Ohmic
dissipation), (2) misalignment between the magnetic and rotation
axes, and (3) turbulence (see \ct{Li+2014} for a critical
review of the proposed mechanisms). Ambipolar diffusion does not
appear to weaken the braking enough to enable large-scale RSD
formation under realistic conditions (\ct{KrasnopolskyKonigl2002};
\ct{MellonLi2009}; \ct{DuffinPudritz2009}; \ct{Li+2011}).
Ohmic dissipation can produce small, AU-scale, RSDs in the early protostellar accretion phase (\ct{Machida+2011}; \ct{DappBasu2010};
\ct{Dapp+2012}; \ct{Tomida+2013}). How such disks grow in time remain
to be fully quantified. Larger, $10^2\au$-scale RSDs can be produced if
the resistivity or the Hall coefficient of the dense core is
much larger than the classical (microscopic) value (\ct{Krasnopolsky+2010};
\ct{Krasnopolsky+2011}; \ct{Santos-Lima+2012}; \ct{BraidingWardle2012a};
\ct{BraidingWardle2012b}). Large RSDs can also form if the magnetic field is misaligned with the rotation axis by a large angle (see \ct{Hull+2013} for observational evidence for misalignment but \ct{Davidson+2011} and \ct{Chapman+2013} for evidence to the contrary), provided that the dense core is not too strongly magnetized (\ct{Joos+2012}; \ct{Li+2013}; \ct{Krumholz+2013}).

The effects of turbulence on magnetized disk formation were first
explored by \citet{Santos-Lima+2012}, who demonstrated that a strong
enough turbulence can enable the formation of a $10^2\au$-scale RSD\@.
The beneficial effects of turbulence on disk formation were confirmed
numerically by a number of authors, including
\citet{Seifried+2012,Seifried+2013},
\citet{Santos-Lima+2013}, \citet{Myers+2013},
and \citet{Joos+2013}. However, why the turbulence is conducive to
disk formation remains hotly debated.
\citet{Santos-Lima+2012,Santos-Lima+2013} attributed
the disk formation to the turbulence induced or enhanced magnetic
reconnection (\ct{LazarianVishniac1999}; \ct{Kowal+2009}),
which reduces the strength of the magnetic field in the inner,
disk-forming, part of the accretion flow.
\citet{Seifried+2012,Seifried+2013} proposed instead
that the turbulence-induced tangling of field lines and strong local
shear are mainly responsible for the disk formation: the disordered
magnetic field weakens the braking and the shear enhances rotation.
\citet{Joos+2013} found that the turbulence produced an effective
magnetic diffusivity that enabled the magnetic flux to diffuse
outward, broadly consistent with the picture envisioned in
\citet{Santos-Lima+2012,Santos-Lima+2013}. It also generated a substantial
misalignment between the rotation axis and magnetic field direction
(an effect also seen in \ct{Seifried+2012} and \ct{Myers+2013}),
which is known to promote disk formation. The lack of consensus on
why turbulence helps disk formation in magnetized cloud cores
motivated us to examine this important issue more closely.

We carry out numerical experiments of disk formation in magnetized dense
cores with different levels of initial turbulence. We find that the
magnetic flux threading the circumstellar, disk-forming region on the
equatorial plane is indeed reduced by turbulence. We show that this
reduction can be explained by a combination of magnetic decoupling-triggered
reconnection of severely pinched field lines close to the central object and a simple geometry effect
--- warping of the well-known magnetic pseudodisk (\ct{GalliShu1993})
out of the disk-forming, equatorial plane by turbulence --- without
having to rely on turbulence-induced magnetic reconnection.
We find that the turbulence-induced pseudodisk warping also reduces
the rates of angular
momentum removal by both magnetic torque and outflow, which is
conducive to disk formation. We describe the problem setup in
\S\ \ref{setup}. The numerical results are presented and analyzed in
\S\ \ref{results}. We compare our results to previous work and discuss
their implications in \S\ \ref{discussion} and conclude with a summary
in \S\ \ref{conclusion}.

\section{Problem Setup}
\label{setup}

We will adopt the basic setup of
\citeauthor{Santos-Lima+2012}\ (\citeyear{Santos-Lima+2012};
see also \ct{Krasnopolsky+2010}),
where a rotating,
magnetized, turbulent but non-self-gravitating dense core accretes
onto a central object of fixed mass. This setup is idealized, but
has an important advantage
for our purpose of understanding why turbulence helps disk formation
in a magnetized core. It enables us to repeat the type of
calculations by \citet{Santos-Lima+2012}, but
at a better spatial
resolution
in the disk-forming region (and a lower numerical
diffusivity for the magnetic field).
The higher resolution is achieved using the ZeusTW code
(\ct{Krasnopolsky+2010}), which can follow the core collapse
and disk formation on a non-uniform grid in a spherical
polar coordinate system. This coordinate system is more natural
than the Cartesian coordinate system for
disk formation simulations,
especially for implementing clean boundary
conditions near the accreting protostar for both the matter and
magnetic field (see \S\ 2.2 of \ct{MellonLi2008} and discussion
below).
Our goal is to develop a qualitative understanding based on simple
numerical experiments. Quantitative results may be modified when
self-gravity is included (see discussion in \S\ \ref{caveat}).

Following \citet{Li+2011} and \citet{Krasnopolsky+2012},
we start our simulations from a uniform,
spherical core of $1\msun$ and radius $R_0=10^{17}\cm$ in a
spherical coordinate system $(r,\theta,\phi)$. The initial
density $\rho_0=4.77 \times 10^{-19}\rhounit$
corresponds to a molecular hydrogen number density of
$10^5\cm^{-3}$. We adopt an
isothermal equation of state with a
sound speed $a=0.2\kms$ below a critical density $\rho_c=10^{-13}\rhounit$,
and a polytropic equation of state $p\propto\rho^{5/3}$
above it. Following \citet{Krasnopolsky+2010}, we adopt the
following prescription for the initial rotation speed:
\begin{equation}
v_\phi=v_{\phi,0} \tanh (\varpi/\varpi_c),
\label{rotation}
\end{equation}
which implies that the equatorial plane is the plane of disk
formation. We adopt $v_{\phi,0}=2\times 10^4\cms$ and
$\varpi_c= 3\times 10^{15}\cm$ to ensure that a large, well resolved,
rotationally supported disk is formed at a relatively early time
in the absence of any magnetic braking (see Fig.\ 1 of
\ct{Krasnopolsky+2010} and Fig.\ \ref{Disk} below). The goal of
our numerical experiments is to determine
whether such a disk is suppressed by a realistic level of magnetic
field in the absence of turbulence and, if yes, whether turbulence
can weaken the magnetic braking enough to allow the disk to
reappear.

Since the focus of our investigation is on the effects of turbulence,
we will consider only one value, $B_0=35.4\muG$, for the strength
of the magnetic field, which is assumed to be uniform initially
and aligned with the rotation axis (i.e., with a misalignment
angle $\theta_0=0^\circ$; a misaligned case of $\theta_0=90^\circ$
will be discussed in \S\ \ref{discussion}). It corresponds to a dimensionless
mass-to-flux ratio $\lambda = 2.92$, in units of the critical value
$(2\pi G^{1/2})^{-1}$, for the core as a whole, which is not far from the
median value of $\lambda\sim 2$ inferred by \citet{TrolandCrutcher2008}
for a sample of nearby dense cores. The mass-to-flux ratio for the
central flux tube $\lambda_c$ is higher than the global value
$\lambda$ by $50\%$, so that $\lambda_c= 4.38$. Our chosen magnetic
field is therefore not unusually strong; if anything, it may be on
the weak side.

We add a turbulent velocity field to the magnetized core at the
beginning of the simulation. Although the existence of ``turbulence''
on the $0.1\pc$ core scale has been known for a long time through
``non-thermal'' linewidth (e.g., \ct{Myers1995}), its detailed
properties, such as energy spectra, remain poorly constrained
observationally.
For simplicity, we generate the initial turbulent velocity field as a
superposition of 1000 sinusoidal waves of wavelengths logarithmically
spaced between a minimum wavelength $l_{\min}=2\times 10^{14}\cm$ and
a maximum $l_{\max}=5\times 10^{16}\cm$. The initial velocity vector
of each sinusoidal wave has an amplitude that is proportional to
$l^p$, a random phase, and a random direction that is perpendicular to
an equally random wave propagation vector.\footnote{Projection from Cartesian to spherical
components on the severely non-uniform grid can distort this
picture somewhat. It can introduce, for example, a small deviation from
the zero-divergence in the initial velocity field.
It is also expected to introduce aliasing of short
wavelength waves inside the coarse resolution regions, which are
located at large radii in our simulations.
In addition, the minimum wavelength resolved in the calculation
changes from place to place (as is also true for other non-uniform grids,
such as in AMR)\@. How this would affect the simulation results
remains to be quantified.}
We have experimented with different number of waves and random
seeds, and found qualitatively similar results. The main parameter
that we decide to vary is the level of turbulence, which is
characterized by the rms Mach number $M$. In \S\ \ref{results},
we will discuss in some depth six models that have the same
turbulent velocity field except for the overall normalization,
which is given respectively by $M=0$ (non-turbulent,
Model A of Table 1), $0.1$ (Model B), $0.3$ (Model C), $0.5$
(Model D), $0.7$ (Model E), and $1$ (Model F)\@. In addition, we will
consider the disk formed in a non-magnetic, non-turbulent core
(Model H) and a case with the magnetic field orthogonal to the
rotation axis ($\theta_0=90^\circ$,
Model P), for comparison with the disks formed in the aligned
($\theta_0=0^\circ$) case that are enabled by turbulence
(see \S\ \ref{discussion}).

\begin{deluxetable}{llllll}
\tablecolumns{6}
\tablecaption{Models \label{table:first}}
\tablehead{
\colhead{Model}
& \colhead{$\lambda^{\rm a}$}
& \colhead{$M$}
& \colhead{$p$}
& \colhead{$\theta_0$}
& \colhead{RSD$^{\rm b}$ }
}
\startdata
A   & 2.92  & 0.0 & N/A & 0$\degree$  & No \\
B   & 2.92  & 0.1 & 1 & 0$\degree$  & No \\
C   & 2.92  & 0.3 & 1 & 0$\degree$  & No \\
D   & 2.92  & 0.5 & 1 & 0$\degree$  & Yes/Transient \\
E   & 2.92  & 0.7 & 1 & 0$\degree$  & Yes/Transient \\
F   & 2.92  & 1.0 & 1 & 0$\degree$  & Yes/Persistent \\

H   & $\infty$ & 0.0 & N/A & 0$\degree$ &  Yes/Persistent \\ 

P   & 2.92  & 0.0 & N/A & 90$\degree$ &  Yes/Persistent \\ 

U   & 2.92  & 1.0 & 0.5 & 0$\degree$  &  Yes/Persistent \\
V   & 2.92  & 1.0 & 2.0 & 0$\degree$  &  Yes/Persistent \\

\enddata
\tablecomments{(a) The average dimensionless mass-to-flux ratio for
the core as a whole. (b) ``Persistent'' disks are rotationally
supported structures that rarely display large deviations from smooth
Keplerian motions, whereas ``transient'' disks are highly active,
rotationally dominated structures with large distortions and are
often completely disrupted.}
\end{deluxetable}

\begin{figure}
\epsscale{1.0}
\plotone{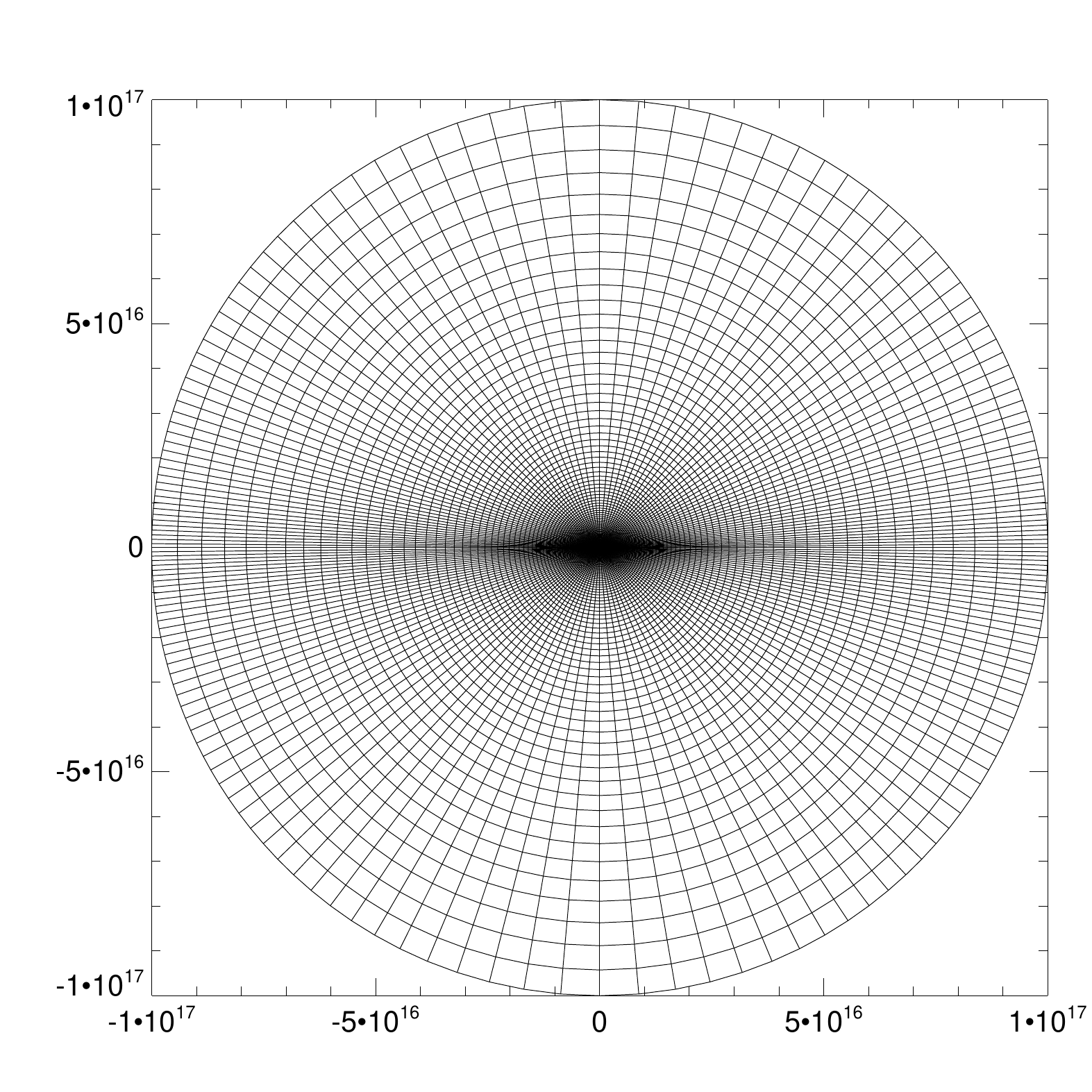}
\caption{Computational grid in the meridian plane, showing good
resolution in the disk-forming equatorial region.
}
\label{grid}
\end{figure}

We choose a non-uniform grid of $120\times 90\times 90$. In the radial
direction, the inner and outer boundaries are located at $r=10^{14}$
and $10^{17}\cm$, respectively. The radial cell size is smallest
near the inner boundary ($5\times 10^{12}\cm$ or $\sim 0.33\au$). It
increases outward by a constant factor $\sim 1.06$ between adjacent
cells. In the polar direction, we choose a
relatively large cell size ($5\degree$) near the polar axes,
to prevent the azimuthal cell size from becoming prohibitively
small in the polar region; it decreases smoothly to a minimum
of $\sim 0.52\degree$
near the equator, where rotationally supported disks may form.
The grid is uniform in the azimuthal direction.
Our finest cell in the disk-forming equatorial region has
dimensions of $0.33$, $0.07$, and $0.47\au$ in the $r$-, $\theta$-, and
$\phi$-direction, respectively. This is comparable to that of
\citet[$0.4\au$]{Joos+2013}, and better than those of
\citet[$1.2\au$]{Seifried+2012}, \citet[$10\au$]{Myers+2013}, and
\citet[$15.6\au$]{Santos-Lima+2012}. The higher resolution should reduce the level
of numerical diffusion of magnetic field and its associated
reconnection, especially compared to that in \citet{Santos-Lima+2012},
whose results we seek to verify and understand physically.
Our non-uniform grid is shown in Fig.\ \ref{grid}. It was designed to
provide good resolution for the disk forming equatorial region. Half of our
radial cells lie within a radius of $\sim 200\au$, and half of our polar
cells within $\sim 20^\circ$ of the equatorial plane. For example, the
relatively thick disk shown in Fig.\ \ref{DiskB} below contains
about $2.5\times 10^5$ cells.

The boundary conditions in the azimuthal direction are periodic. In
the radial direction, we impose the standard outflow boundary
conditions for both the hydrodynamic quantities and magnetic field,
at both the inner and outer boundaries.
The boundary conditions are enforced using ghost zones. For the
density and three components of the velocity, we simply copy their
values in the active zone closest to the boundary into the ghost
zones, except when the radial component of the velocity is pointing
into the computation domain (i.e., $v_r > 0$ near the inner boundary
or $v_r < 0$ near the outer boundary); in such cases, the radial
velocity is set to zero in the ghost zones to prevent mass flowing
into the computation domain from outside, where there is no
self-consistently determined flow information. These boundary
conditions allow matter and angular momentum to leave the outer radial
boundary, as needed for the magnetic-braking driven outflow, and to
accrete through the inner radial boundary. The boundary conditions on
the magnetic field are enforced through the electromotive force (EMF),
which is used to evolve the field everywhere, including the ghost
zones, using the method of constrained transport (CT)\@. The three
components of the EMF are copied from the active zone closest to the
boundary into the ghost zones. In effect, we are assuming continuity
from the computation domain into the ghost zones for both the
hydrodynamic quantities and magnetic field, which minimizes the risks
of artificial wave reflection at the boundary. The use of CT
ensures that the divergence-free condition $\nabla\cdot {\bf B}=0$ is
preserved to machine accuracy in both the active computation domain
and the ghost zones. In particular, the magnetic field lines dragged
by the accretion flow across the inner boundary stay on the boundary,
forming essentially a split magnetic monopole, as expected in
ideal MHD (\ct{Galli+2006}), until (numerical) reconnection
is triggered by severe pinching of the oppositely directed field
lines (see \ct{MellonLi2008} and discussion below). The radius of our
inner boundary is $10^{14}\cm$ (or $6.7\au$). It is larger than the sink
accretion radius used by \citet[$3\au$]{Seifried+2013} but smaller
than that of \citet[$62.5\au$]{Santos-Lima+2012}. Since our inner
boundary is covered by nearly 10,000 cells, it can resolve the angular
distributions of the magnetic field and flow quantities better than the
``sink accretion region'' of \citet{Santos-Lima+2012} and
\citet{Seifried+2013}. We note that there was formally no ``sink accretion
region'' in \citet{Joos+2013}. They adopted a stiffened equation of
state above a mass density of $10^{-13}\rhounit$, which produced
an artificially thermally supported object of order $10\au$ in size
(comparable to that of our inner boundary), which served as an effective ``inner
boundary'' for their simulations. Note that their ``inner boundary''
is quite different from the ``sink accretion region'' of
\citet{Santos-Lima+2012}
and \citet{Seifried+2013}, and both treatments are very
different from ours. How these different treatments affect the
numerical results remains to be quantified.

On the polar axes, the boundary condition is chosen to be reflective.
Although this is not strictly valid, we expect its effect to be
limited to a small region near the axis. As in
\citet{Krasnopolsky+2010} and \citet{Santos-Lima+2012}, the central
point mass is fixed at $0.5\msun$.

Although it is desirable to carry out the simulations in ideal
MHD, so that they can be compared more directly with other
works, especially \citet{Santos-Lima+2012}, we found the ideal
MHD simulations difficult
to perform in practice, because they tend to produce numerical ``hot zones''
where the Courant conditions demand such a small timestep that they force the calculation to stop early, a tendency we noted in
our previous 2D (\ct{MellonLi2008}) and 3D simulations
(\ct{Krasnopolsky+2012} and \ct{Li+2013}).
To lengthen the simulation, we include a small, spatially
uniform resistivity $\eta=10^{17}\etaunit$. We have
verified that, in Model F with $M=1$ (which turns out to be one of the most
interesting cases and will be discussed in greatest detail),
this resistivity changes the flow structure little compared to
either the ideal MHD limit or a model where the resistivity is
reduced by a factor 10, to $10^{16}\etaunit$, at early times
(before the non-resistive and low-resistivity runs stop). It is at
least two orders of magnitude smaller than the value needed to
enable large-scale disk formation by itself (\ct{Krasnopolsky+2010}).

There was no mention of the need for using explicit resistivity to lengthen
simulation of magnetized disk formation by other groups. The exact
reason for this difference is unclear, although we suspect that it is
related to the relatively low magnetic diffusivity in our code from
the use of (1) fixed non-uniform grid, which avoids the numerical
diffusion associated with refinement and derefinement, and (2) method
of characteristics in constrained transport, which makes the MHD
algorithm more accurate (\ct{StoneNorman1992}).
The results of the current simulations from different groups appear to
depend strongly on the numerical code used in each study.
In the
future, it will be desirable to benchmark all MHD codes used for disk
formation simulations against common test problems such as the
collapse of a non-rotating, uniformly magnetized sphere of constant
density, especially in the protostellar accretion phase, when the
magnetic field is severely pinched and its structure is sensitive to
the level of numerical diffusivity (see related discussion in
footnote \ref{2D} below).

\section{Numerical Results and Analysis}
\label{results}

\subsection{Turbulence-Enabled Disk Formation}
\label{DiskFormation}

\begin{figure}
\epsscale{1.0}
\plotone{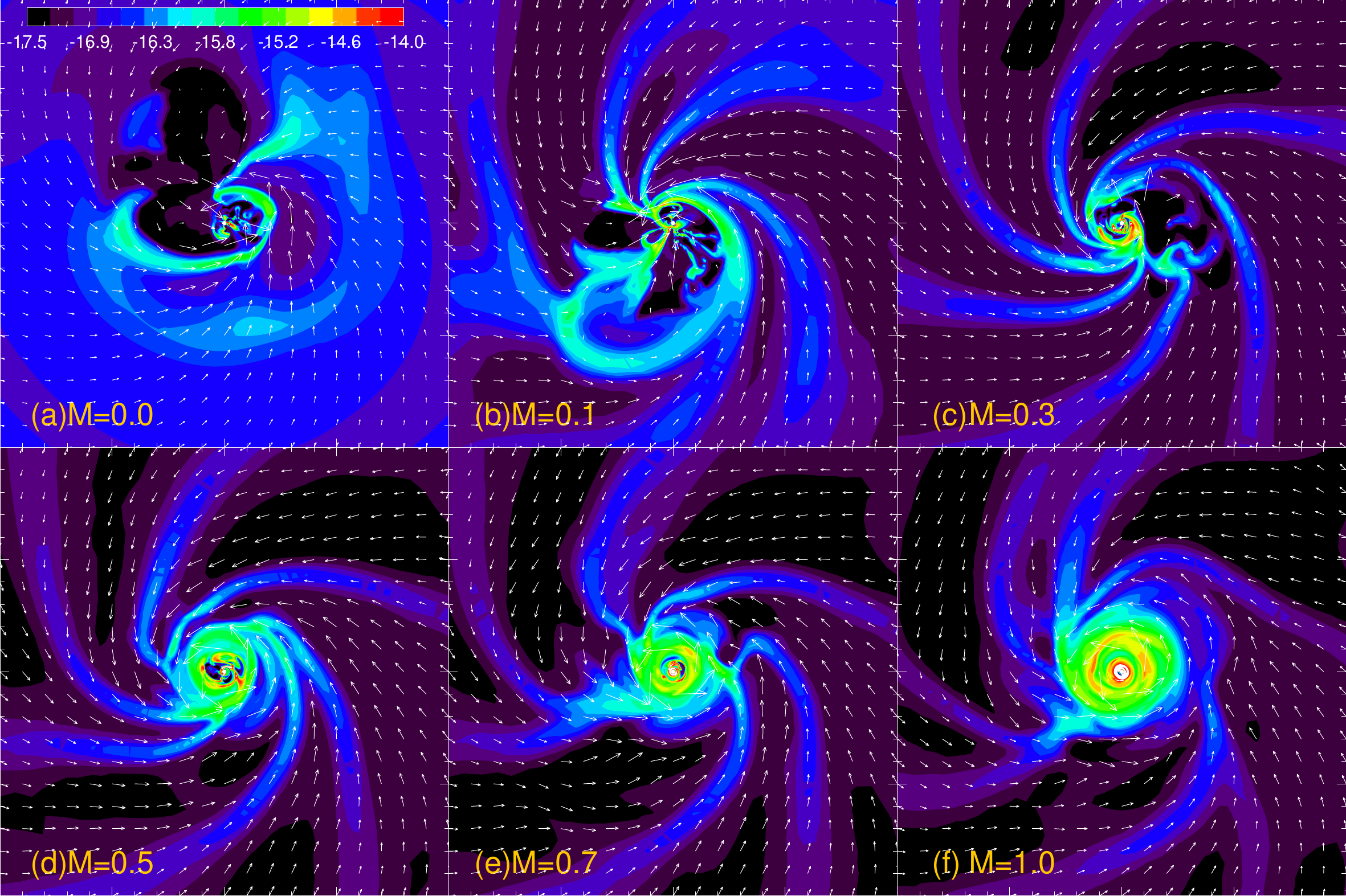}
\caption{Density distribution and velocity field on the equatorial
plane at a representative time ($t=8\times 10^{11}\second$) for models
with different levels of turbulence (Model A--F)\@. The beneficial
effect of
turbulence on disk
formation is evident. Each panel is $2\times 10^{16}\cm$ on the
side. The color bar is for the logarithm of the density added to a floor value of
$10^{-17.5}\rhounit$. Equatorial values taken at a latitude
between $0$ and $0.26\degree$.
}

\label{TurbE_6p}
\end{figure}

To illustrate the effects of turbulence on disk formation in
magnetized dense cores, we carried out a set of six simulations with
identical initial conditions except for the level of turbulence,
which is characterized, respectively, by the rms turbulent Mach
number $M=0$, $0.1$, $0.3$, $0.5$, $0.7$, and $1$ (see Models A--F in
Table 1). The simulations were run to a common final time
$t=1.1\times 10^{12}\second$ or about $3.5\times 10^4$ years.
Fig.\ \ref{TurbE_6p} shows the density distribution and velocity
field on the equatorial plane at a representative time
$t=8\times 10^{11}\second$ for all six cases. The difference in
morphology is striking.

In the non-turbulent ($M=0$) model (Model A), there is no
hint of any rotationally supported disk (RSD), consistent with
previous work. The circumstellar
region is dominated by a highly magnetized, low-density expanding
region (the so-called ``decoupling enabled magnetic structure'' or
DEMS that has been discussed extensively in \ct{Zhao+2011} and
\ct{Krasnopolsky+2012}; see Fig.\ \ref{DEMS} and
\S\ \ref{dems_discussion} below). This behavior
is not changed fundamentally by a modest amount
of turbulence in the $M=0.1$ (Model B) or $0.3$ (Model C) cases,
where the RSD remains suppressed and DEMS remains dynamically
important. Here, the turbulence does change the appearance
of the density distribution on the equatorial plane drastically.
It produces well-ordered dense spirals that are absent in the
non-turbulent case. The spirals mark the
locations where a thin, warped, pseudodisk (shown in Figs.\ \ref{DenSphere} and
\ref{3DView} below) intercepts the
equatorial plane.

\begin{figure}
\epsscale{1.0}
\plotone{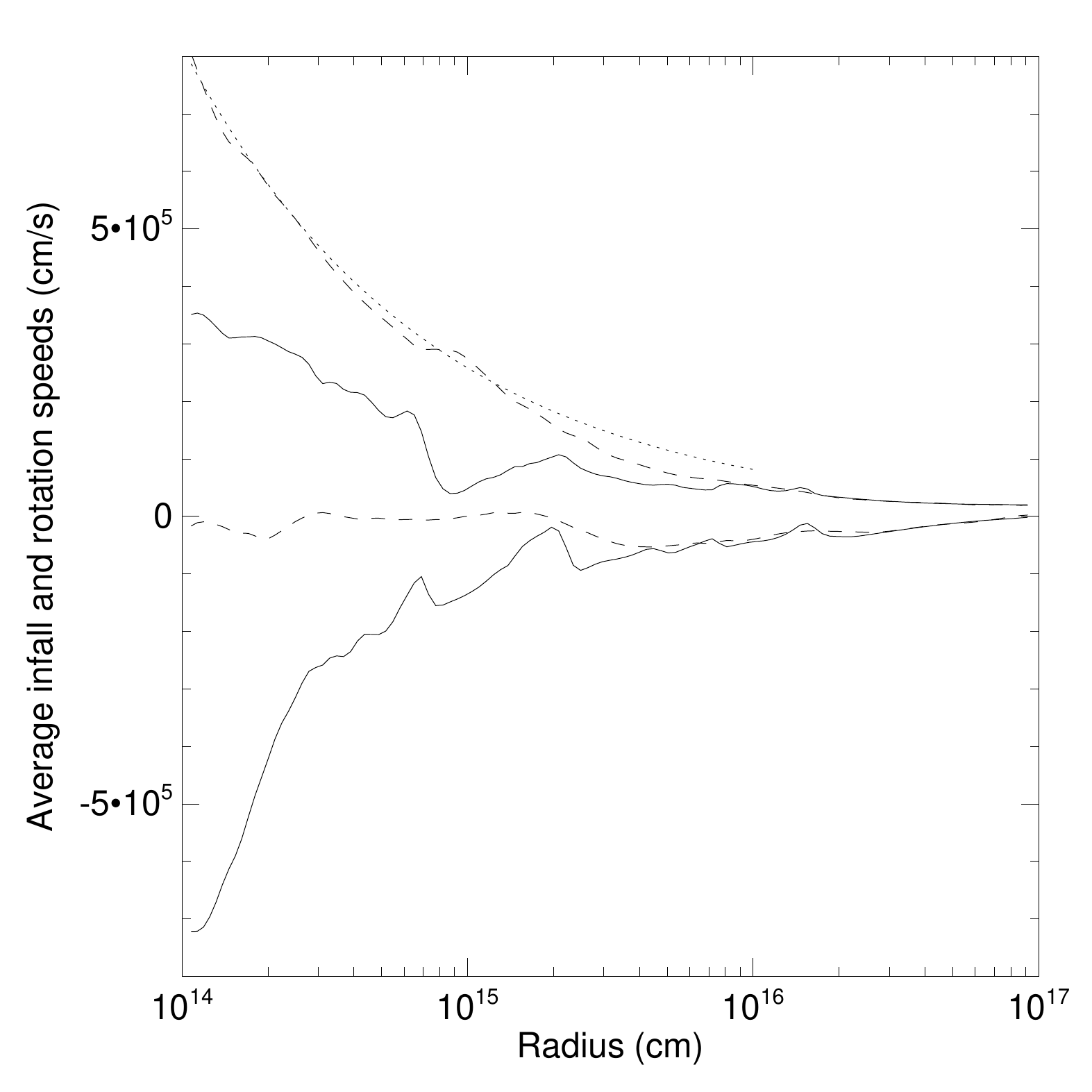}
\caption{Mass-weighted infall (lower two curves) and rotation (upper
two) speed as a function of radius for the non-turbulent (Model A,
solid curves) and sonic turbulence (Model F, dashed) cases. The
averaging is done within $20^\circ$ of the equatorial plane. A
Keplerian profile (dotted line) is shown for comparison.
}
\label{InfallRot}
\end{figure}

The apparent spirals persist as the level of turbulence increases.
At the time shown in Fig.\ \ref{TurbE_6p}, they appear to
merge into a disk-like structure in Model D ($M=0.5$), although the
central region of the structure is still filled with low-density
``holes.'' This porous disk is highly dynamic. It forms around $\sim
6\times 10^{11}\second$, and becomes disrupted by $\sim 9\times
10^{11}\second$ (a movie illustrating the transient nature of the
disk is available online as auxiliary material). The situation is similar
in the slightly stronger turbulence case of $M=0.7$ (Model E),
where a transient disk is
also formed. Compared to the $M=0.5$ case, the disk in the $M=0.7$
case forms earlier ($\sim 2\times 10^{11}\second$), and lasts longer
(until $\sim 9.5\times 10^{11}\second$). As the level of turbulence
increases to $M=1$ (Model F), a well-defined disk forms earlier
still ($\sim 10^{11}\second$), grows steadily with time, and persists
to the end of the simulation.
As seen from Fig.\ \ref{InfallRot}, the disk is rotationally supported, with an
average rotation speed close to the Keplerian speed and a much smaller
infall speed inside a radius of $\sim 2\times 10^{15}\cm$ at the time
shown in Fig.\ \ref{TurbE_6p} ($t=8\times 10^{11}\second$). This is in
contrast with the non-turbulent case (Model A) where the rotation on
the same $100\au$ scale is highly sub-Keplerian and is dominated by infall.
The rotationally supported disk in Model F turns out to be rather
thick and significantly magnetized. Its properties will be discussed
in more detail in \S\ \ref{discussion}. Here we focus on the
unmistakable trend that a stronger turbulence leads to the formation
of a more robust disk. The question is: why is the disk formation
suppressed in the non-turbulent or weakly turbulent cases but enabled
by a stronger turbulence?

\begin{figure}
\epsscale{1.0}
\plotone{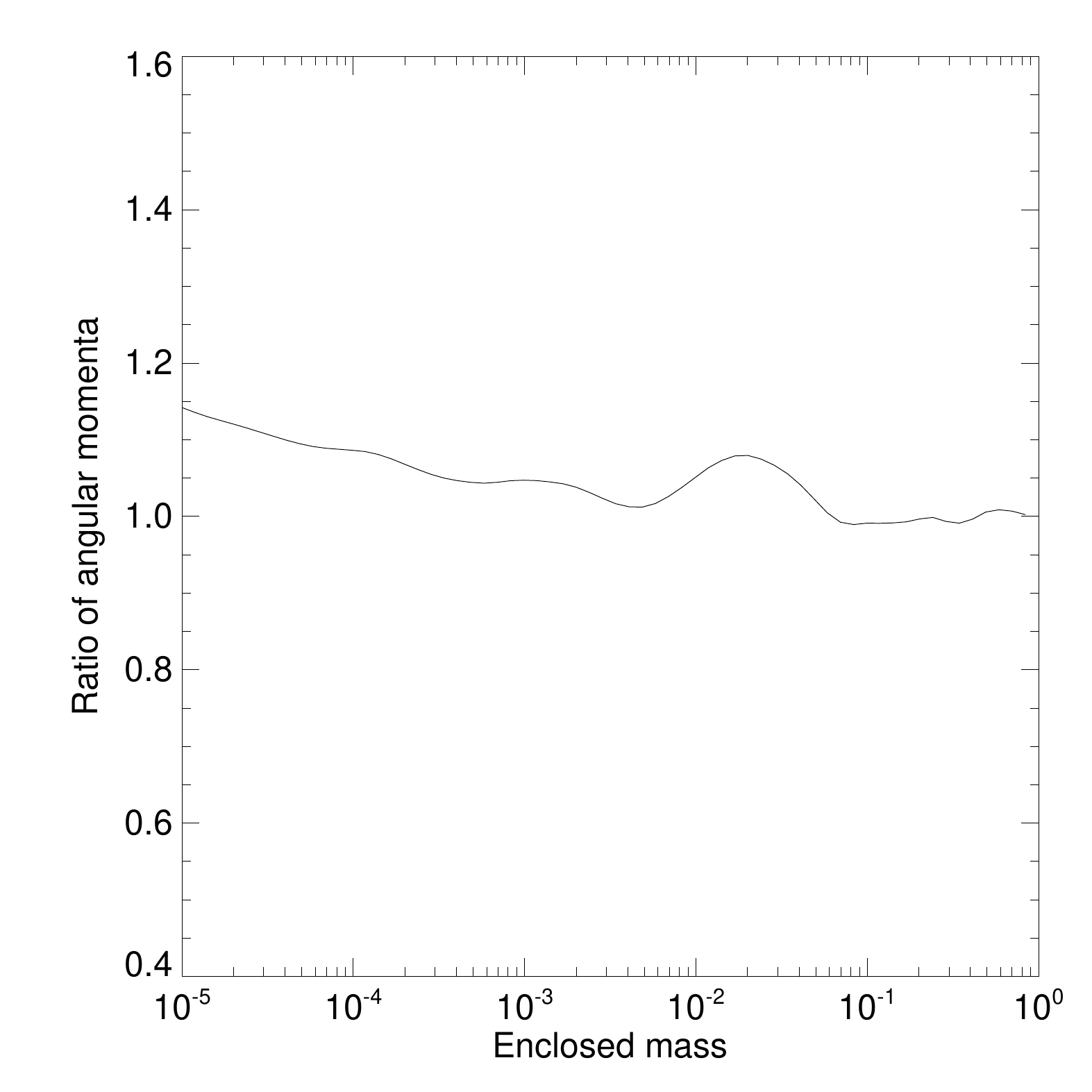}
\caption{The ratio of the initial angular momenta enclosed within a
sphere as a function of the mass enclosed within the same sphere for the
sonic turbulence (Model F) and non-turbulent (Model A) cases,
showing that in these cases the turbulence increases the initial
angular momentum by about $10\%$ or less over most of the mass (and
volume). The mass is normalized to $1\msun$.
}
\label{TurbAngMom}
\end{figure}

One possibility is that a stronger turbulence may increase the initial
angular momentum by a larger amount, making it easier to form a
RSD\@. However, even in the most turbulent case of Model F, the
increase is modest, by $\sim 10\%$ or less over most of the mass
(and volume; see Fig.\ \ref{TurbAngMom}). It is unlikely that such a
modest change alone can explain the drastically different outcomes for our
non-turbulent and sonic turbulence cases.

\subsection{Obstacle to Disk Formation: Central Split Magnetic Monopole}
\label{MagFlux}

Disk suppression in ideal MHD is conceptually tied to
another fundamental problem in star
formation --- the so-called ``magnetic flux problem.'' If the
field lines in a dense core magnetized to the observed level
are dragged by collapse into the central stellar object, they
would produce a split magnetic monopole near the center that is
strong enough to remove essentially all of the angular momentum
of the accreted material and prevent the formation of a rotationally supported
circumstellar disk (\ct{Galli+2006}). However,
it is well known that if the flux freezing holds strictly during
the core collapse, the stellar field strength would be orders
of magnitude above the observed values
(\ct{Shu+1987}).
This magnetic flux problem must be resolved one way or another,
and its resolution is a prerequisite for disk formation.

\begin{figure}
\epsscale{0.8}
\plotone{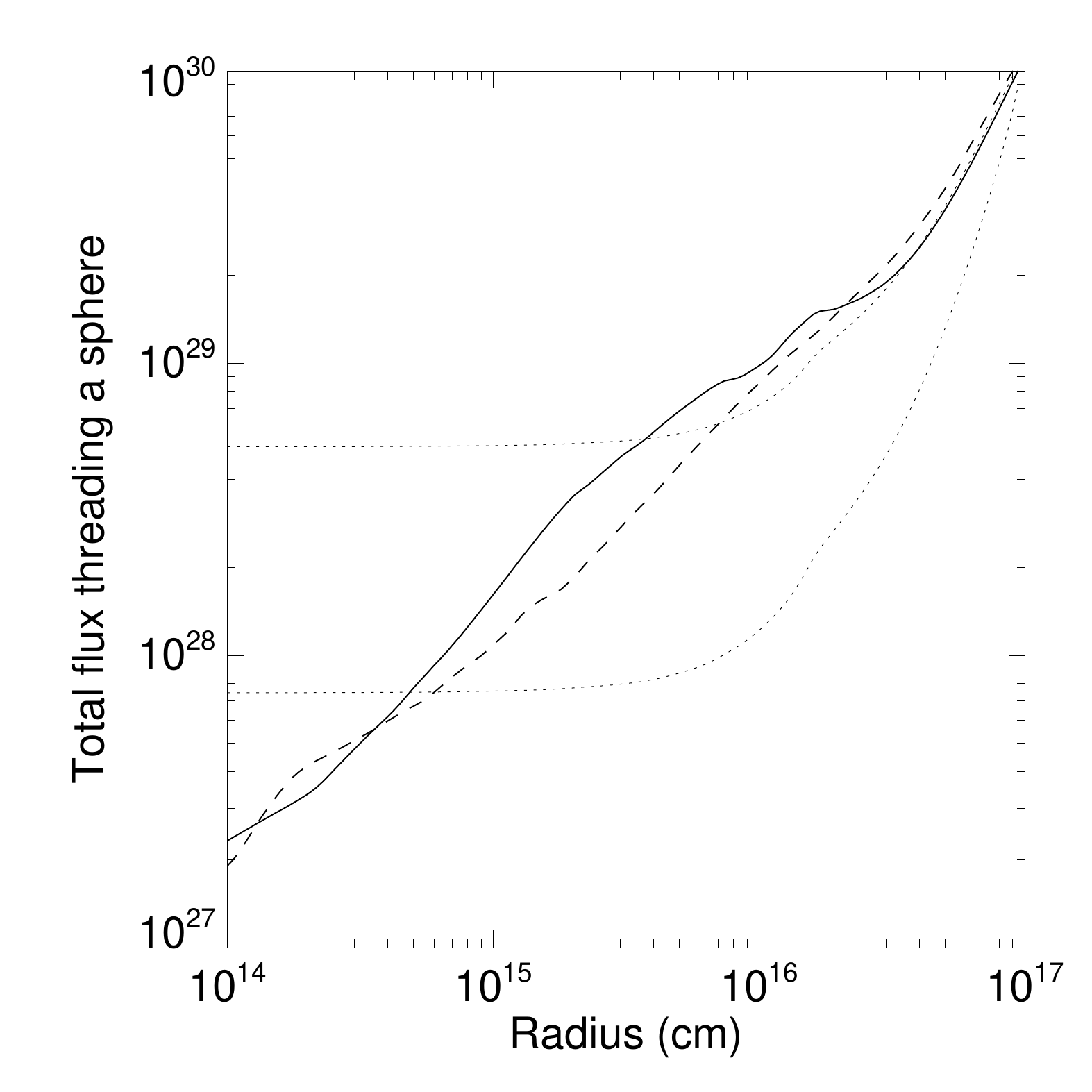}
\caption{Total outgoing magnetic flux $\Phi_r$ threading a sphere (defined in
equation \ref{RadialFlux}, in units of $\gaussnosp\cm^2$) as a function of the
radius of the sphere for the non-turbulent case (Model A, $M=0$,
solid line) and sonic turbulence case (Model F, $M=1$, dashed) at time
$t=8\times 10^{11}\second$ (same as Fig.\ \ref{TurbE_6p}).
Also plotted for comparison are an upper (top dotted line) and lower
(bottom dotted) limit to the magnetic flux expected under the flux
freezing condition, given by
equations (\ref{upper}) and (\ref{lower}), respectively, for Model A\@.
}
\label{diffusion}
\end{figure}

The stellar magnetic flux problem can be resolved in principle
through non-ideal MHD effects
(e.g., \ct{LiMcKee1996}; \ct{Contopoulos+1998}; \ct{KunzMouschovias2010};
\ct{Machida+2011}; \ct{DappBasu2010}; \ct{Dapp+2012}; \ct{Tomida+2013}),
which decouple the field lines from the matter at high densities close
to the central object. In ideal MHD simulations, the decoupling
can be mimicked by numerically induced magnetic flux
redistribution. To demonstrate that flux redistribution
has indeed occurred in our simulations, we plot in Fig.\ \ref{diffusion} the magnetic flux $\Phi_r$
that leaves the surface of a sphere of radius $r$:
\begin{equation}
\Phi_r(r)=\int B_r (>0) \ dS
\label{RadialFlux}
\end{equation}
where the integration is over the part of the surface with
magnetic field pointing outward, i.e., $B_r > 0$ (we have
verified that the amount of flux entering the sphere is exactly
the same as that going out). Near the inner boundary $r_i=10^{14}\cm$,
this flux provides a measure of the strength of the central split
magnetic monopole. Its value on any sphere is to be compared with
the magnetic flux expected to be dragged into the same sphere by
matter under the condition of flux freezing, $\Phi_{r,\rm ff}$. The
expected flux depends on the amount of mass that has
already accumulated inside the sphere ($M(r)$, including the mass that
has passed through the inner boundary),
and whether the mass is accumulated along or across the
field lines; mass accumulation along field lines would not lead
to any flux increase. In the limit that all of the mass
along the field lines that initially thread a sphere has accumulated
inside the sphere, the expected flux would be
\begin{equation}
\Phi_{r,\rm ff}^{\min}(r)=\pi B_0 R_0^2
\left\{
1-\left[1-\frac{M(r)}{M_{\rm tot}}\right]^{2/3}
\right\},
\label{lower}
\end{equation}
where $B_0$, $R_0$ and $M_{\rm tot}=\frac{4\pi}{3}\rho_0 R_0^3$ are the
initial field strength, radius, and total mass of the core. This
flux is a (generous) lower limit to $\Phi_{r,\rm ff}$ at relatively
small radii, where only a small fraction of the mass along any
given field line has collapsed close to the central object at the
relatively early times under consideration. It is derived by
relating the magnetic flux $\Phi$ enclosed within a cylinder of
radius $\varpi$ in the initially constant-density dense core with
a uniform magnetic field to the mass enclosed by the same cylinder.
An upper limit to
$\Phi_{r,\rm ff}$ is obtained by assuming that the mass accumulation
is isotropic, which yields
\begin{equation}
\Phi_{r,\rm ff}^{\max}(r)=
\pi B_0 R_0^2 \left[\frac{M(r)}{M_{\rm tot}}\right]^{2/3}.
\label{upper}
\end{equation}
This is an upper limit because the core collapse proceeds
somewhat faster along field lines than across.

\begin{figure}
\epsscale{1.0}
\plotone{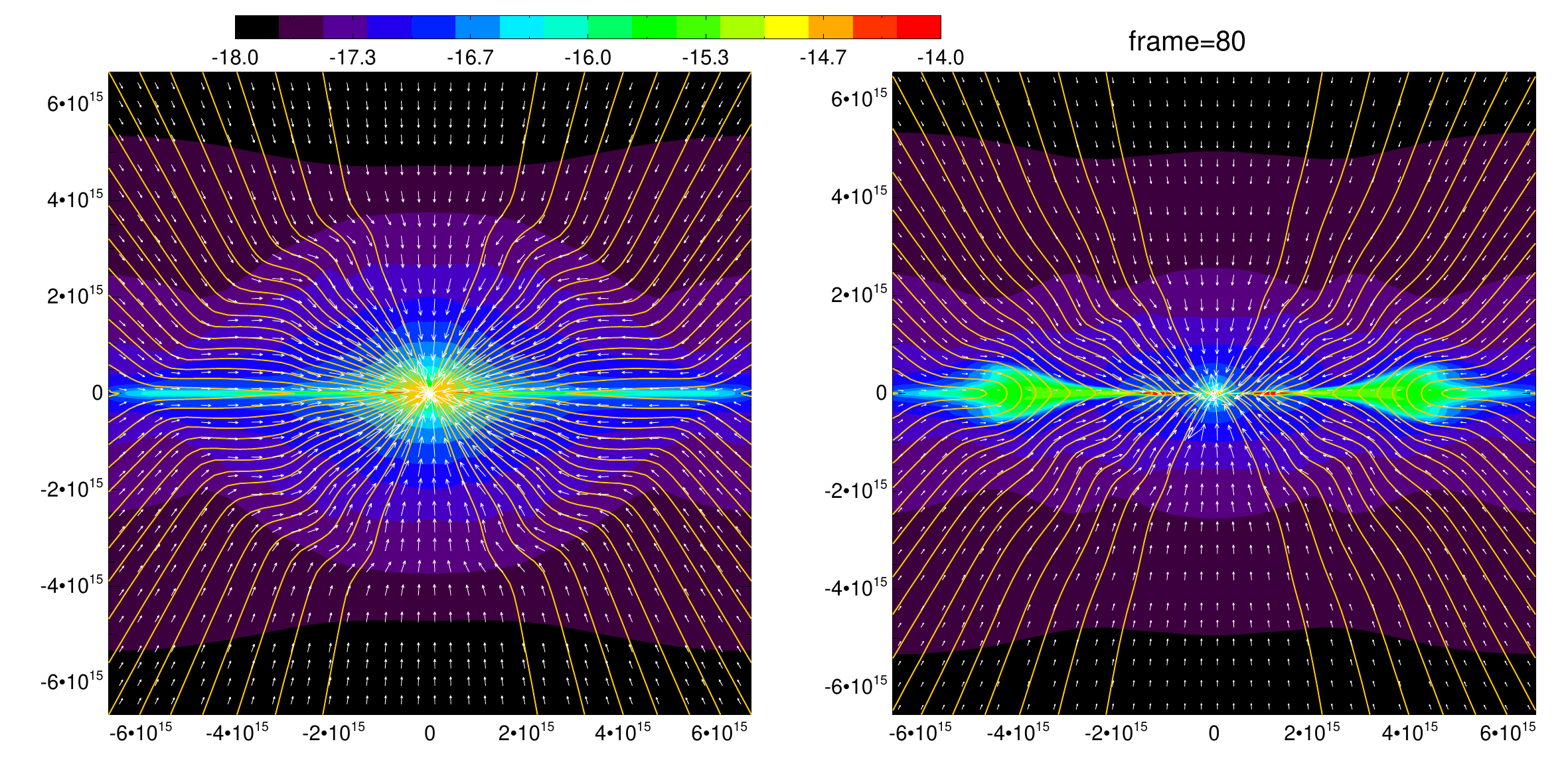}
\caption{Snapshots of two 2D (axisymmetric) simulations with different
  inner boundary radii ($r_i=10^{14}\cm$ for the left panel, and
  $1.5\times 10^{13}\cm$ for the right), showing the magnetic field
  lines (yellow lines), velocity vectors (white arrows) and the
  logarithm of density (color map) in the meridian plane. Note the severely pinched field
  lines before (episodic) reconnection in the left panel, and the two high
  density equatorial ``blobs'' created by (episodic) reconnection in the right
  panel. See movies online.
 }
\label{2DSnap}
\end{figure}

From Fig.\ \ref{diffusion}, it is clear that the actual magnetic flux
$\Phi_r$ is below the minimum value $\Phi_{r,\rm ff}^{\min}$ expected
from flux-freezing at small radii ($r\lesssim 4\times 10^{14}\cm$).
This is evidence for magnetic flux redistribution, which has reduced
the flux near the inner boundary (and hence the strength of the split
magnetic monopole) by at least a factor of 4 (more likely
an order of magnitude) in the non-turbulent case (Model A,
solid line in the figure). The flux redistribution is likely
related to a similar behavior that \citet{MellonLi2008} observed
in their 2D (axisymmetric)
self-gravitating simulations. They found episodic reconnection of
the oppositely
directed field lines above and below the equatorial plane near the
inner boundary (see their Fig.\ 16).
We have carried out several 2D (axisymmetric) non-self-gravitating
simulations with different spatial resolutions and different values of
resistivity $\eta$ (including $\eta=0$), and found episodic
reconnection in all cases. Movies of two examples are included as
online auxiliary material, and their snapshots at a representative
time $t=8\times 10^{11}\second$ (or frame 80) are shown in Fig.\ \ref{2DSnap}. They have the same initial mass and
magnetic field distributions as Models A--F but with $\eta=0$ and
without any initial rotation, and have inner radius $r_i=10^{14}$ and $1.5\times
10^{13}\cm$, respectively.\footnote{Although episodic
reconnection dominates the accretion dynamics close to the central
object in both cases, individual reconnection events can look rather
different (see Fig.\ \ref{2DSnap}). This is perhaps not too surprising, since there is no
explicit resistivity in these simulations, so the reconnection of
the sharply pinched magnetic field must involve numerical
diffusion. It can occur at different locations (and
with different intensities), depending on the inner radius and
spatial resolution. Such a dependence makes it difficult to obtain
numerically converged solutions, at least (perhaps especially) in the
2D case in the ideal MHD limit. Whether non-ideal MHD effects can
alleviate this difficulty or not remains to be determined.\label{2D}}

\begin{figure}
\epsscale{1.0}
\plotone{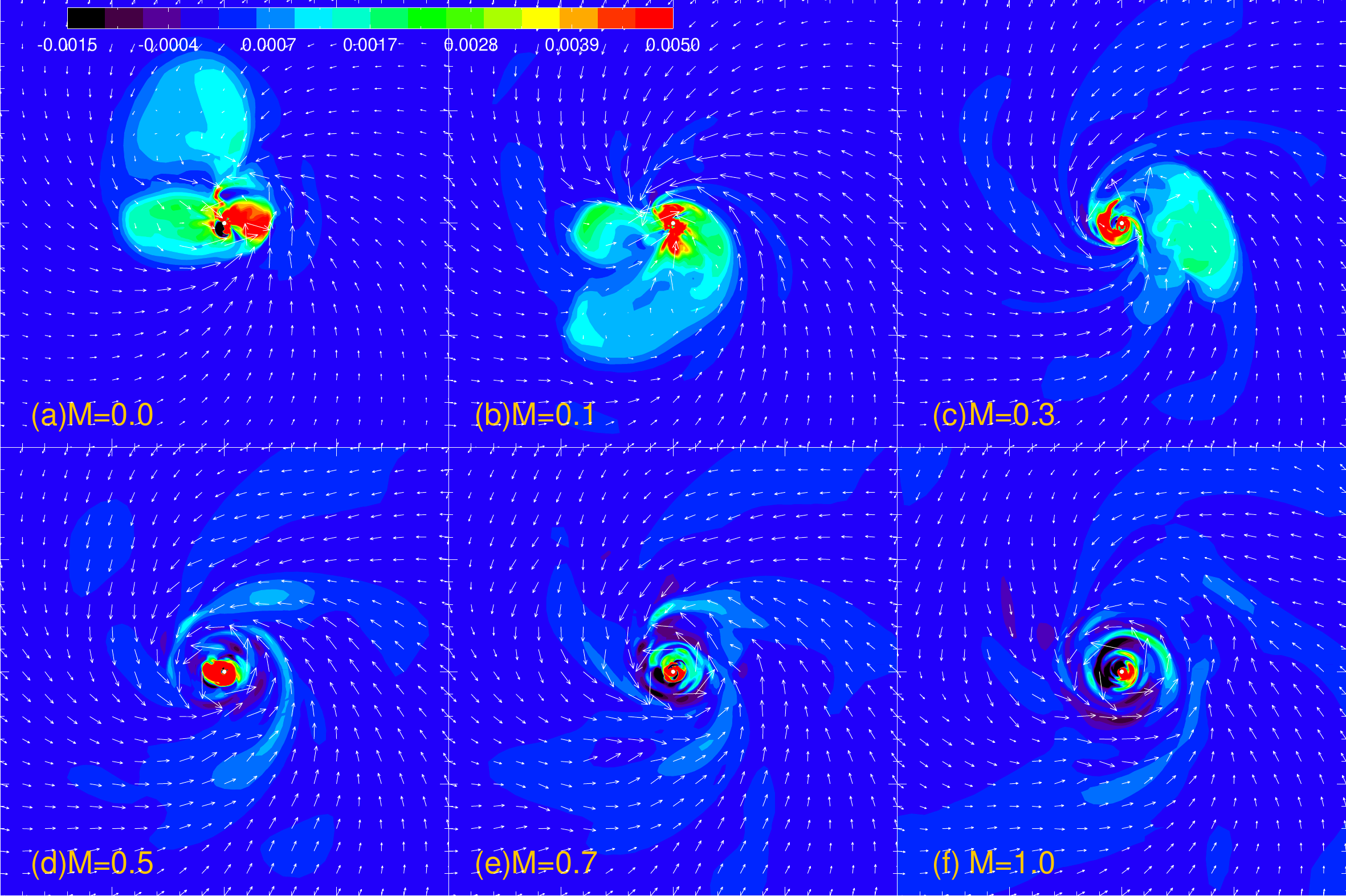}
\caption{Same as in Fig.\ \ref{TurbE_6p}, except for the color map,
which displays the vertical magnetic field strength $B_z$ (in units
of Gauss) on the equatorial plane. Note the strong anti-correlation
between the strongly magnetized region (DEMS, in Models A--C) and
rotationally supported disk (in Models D--F)\@. The weakening of DEMS
appears to be a prerequisite for disk formation.
}
\label{DEMS}
\end{figure}

In any case, the reconnected field lines in the 2D simulations are
driven outward by magnetic tension force, leaving behind a strongly
magnetized, low density region. This two-step flux redistribution
--- field line reconnection near the inner boundary followed by
outward field advection --- is likely operating in our 3D simulations
as well. A difference is that, in 3D, mass accretion can continuously
drag field lines across the inner boundary along some (azimuthal)
directions, with the reconnected field lines escaping
outward along other directions. This more continuous reconnection
makes individual events less powerful, and thus harder to identify
(\ct{Zhao+2011}; \ct{Krasnopolsky+2012}), especially in the presence
of a turbulence. In the non-turbulent (Model A),
and weakly turbulent (Model B and C) cases, the redistributed magnetic
flux remains trapped close to the central object, forming a strongly
magnetized, low-density region --- the DEMS --- that is known to be
a formidable obstacle to disk formation (\ct{Zhao+2011};
\ct{Krasnopolsky+2012}; see Fig.\ \ref{DEMS}). In these cases,
the decoupling-triggered reconnection has greatly weakened the
central split magnetic monopole, which lies at the heart of the
magnetic braking
catastrophe in ideal MHD (\ct{Galli+2006}).
However, it created another, perhaps even more severe, problem ---
the DEMS --- which has to be overcome in order for rotationally
supported disks to form.

\subsection{Obstacle to Disk Formation: DEMS}
\label{dems_discussion}

In order for RSDs to form, both the central split magnetic monopole
{\it and} the DEMS must be weakened. Fig.\ \ref{diffusion} shows that the
amount of magnetic flux $\Phi_r$
threading the inner boundary is about the same for the non-turbulent
(Model A) and sonic turbulence (Model F) cases, indicating that the
weakening of the split magnetic monopole is not controlled by turbulence.
As discussed above, it is most likely caused by the
decoupling-triggered reconnection observed in the 2D axisymmetric
case. The magnetic flux $\Phi_r$ is somewhat lower in the
turbulent case between $\sim 10^{15}\cm$ and $\sim 10^{16}\cm$ (see the
dashed line in Fig.\ \ref{diffusion}). This
could be due to additional, turbulence-enhanced, magnetic reconnection
during the core collapse,
as advocated by \citeauthor{Santos-Lima+2012}\ (\citeyear{Santos-Lima+2012},
see also \ct{Santos-Lima+2013} and \ct{Joos+2013}).
Alternatively, it could be related to how the field lines reconnected
near the inner boundary escape to large distances, as discussed below
in \S\ \ref{origin}. In any case, the difference in $\Phi_r$ between
these models with and without turbulence is relatively moderate.

A more striking difference lies in the DEMS\@. Fig.\ \ref{DEMS},
which plots the vertical field strength $B_z$ on the equatorial
plane, shows that the strongly magnetized DEMS
dominates the circumstellar region on the disk-forming, $10^2\au$
scale for the non-turbulent and weakly turbulent cases (Models A--C)\@.
It becomes much less
prominent for the stronger turbulence cases (Models D--F)\@. The turbulence has
clearly reduced $B_z$ on the equatorial plane (and thus the magnetic flux
threading vertically through the plane) near the central object. This
reduction may hold the key
to understanding the turbulence-enabled disk formation observed in
Fig.\ \ref{TurbE_6p},
given the strong anti-correlation between the highly magnetized DEMS
and the rotationally supported disk.
To quantify the $B_z$ reduction, we focus on the net magnetic
flux $\Phi_z$
that passes vertically through the equatorial plane inside a circle of
cylindrical radius $\varpi$:
\begin{equation}
\Phi_z(\varpi)=\Phi_i+\int_0^{2\pi}\int_{\varpi_i}^\varpi B_z(\varpi,
\theta=\pi/2,\phi) \varpi\ d \varpi\ d\phi\ ,
\label{VerticalFlux}
\end{equation}
where $\Phi_i$ is the contribution from the upper hemisphere of the inner (spherical)
boundary and $\varpi_i=10^{14}\cm$.

\begin{figure}
\epsscale{1.0}
\plottwo{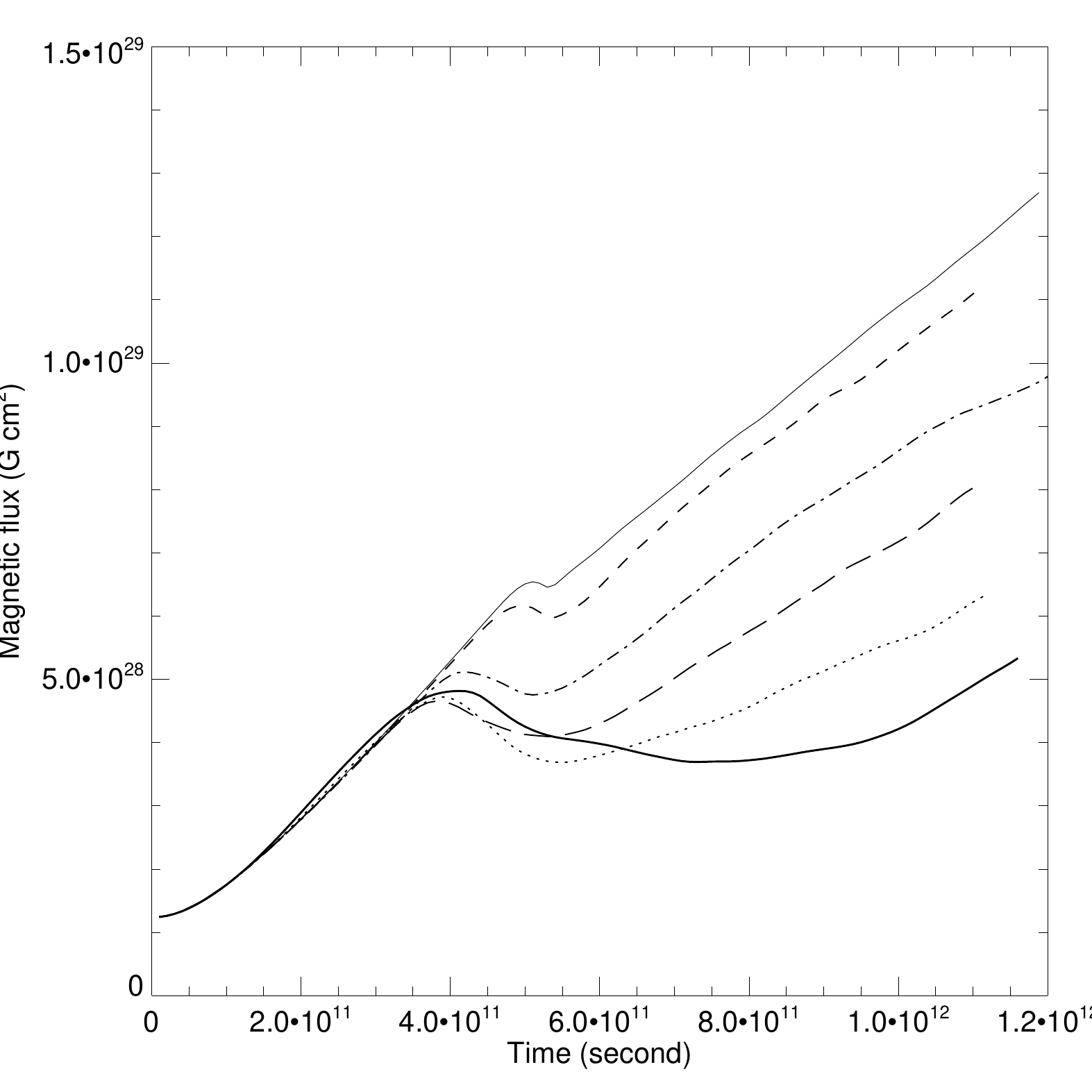}{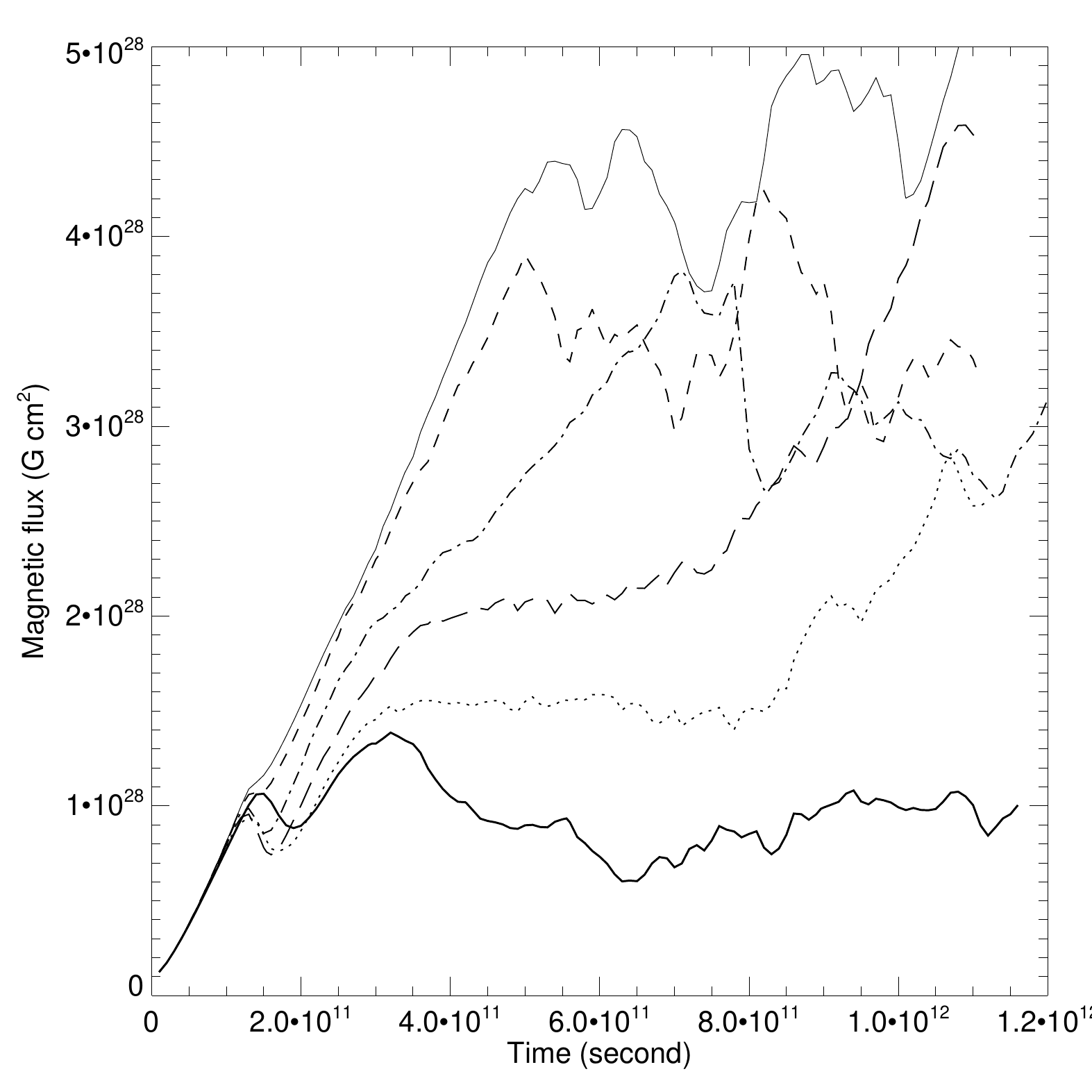}
\caption{Time evolution of the magnetic flux $\Phi_z$ passing
vertically through a circle
on the equatorial plane with cylindrical radius $\varpi=
1.055 \times 10^{16}\cm$ (left panel)
and $3.055 \times 10^{15}\cm$ (right panel) for Model A ($M=0$, upper thin
solid line), B (dashed), C (dash-dotted), D (long dashed), E (dotted)
and F ($M=1$, lower thick solid line). The trend is clear that a stronger
turbulence leads to a lower magnetic flux at late times.
}
\label{Bflux}
\end{figure}

In Fig.\ \ref{Bflux}, we plot the time evolution of the vertical magnetic
flux $\Phi_z$ inside a circle of
cylindrical radius $\varpi \approx 10^{16}\cm$ (left panel)
and $3 \times 10^{15}\cm$ (right panel) for
Models A--F.\footnote{We have verified that $\Phi_z$ is
equal to the magnetic flux that enters the lower hemisphere of a
sphere of radius $r=\varpi$ and that leaves the upper hemisphere of the
same sphere to machine accuracy, as expected for the treatment of
the induction equation using constrained transport.\label{fn1}}
For the larger circle, $\Phi_z$ increases more or less monotonically
with time in the absence of any turbulence ($M=0$, the upper solid line
in the figure). This is to be expected, since more and more field lines
are dragged across the circle as the equatorial material collapses. The
pause around $t_{k}\sim 5\times
10^{11}\second$ is caused by the outer edge of the dense, equatorial
pseudodisk (\ct{GalliShu1993}; see Fig.\ \ref{Pseudo} below for an
example) expanding across the circle under consideration; the pseudodisk
expansion temporarily lowers the flux $\Phi_z$. A similar (although
weaker) kink is also visible for the smaller, $\varpi\approx 3\times 10^{15}\cm$,
circle (see the right panel of Fig.\ \ref{Bflux}). It occurs at an
earlier time $t_{k}\sim 10^{11}\second$, which is expected since the
outer edge of the pseudodisk crosses this smaller circle sooner. The
most striking feature for the non-turbulent case is that the magnetic
flux inside the smaller circle levels off after $t\sim 5\times
10^{11}\second$, and becomes oscillatory. The oscillation is caused by
the highly magnetized, low-density structure (DEMS, see
Figs.\ \ref{TurbE_6p} and \ref{DEMS}) expanding beyond the circle,
advecting back out the magnetic flux dragged across the circle by the
collapsing flow in a highly time variable manner.

In the presence of a turbulence, the time evolution of the magnetic
flux $\Phi_z$ changes significantly. For the larger circle ($\varpi
\approx 10^{16}\cm$), as the level of turbulence increases, there is a
trend for the kink on the $\Phi_z(t)$ curve to start earlier,
the turnover to last longer, and the increase after the turnover
to become slower. The earlier kink occurs because the outer
edge of the pseudodisk is distorted by turbulence, causing it
to reach the circle earlier. In the strongest turbulence case
(Model F with $M=1$), the magnetic flux stays more or less
constant after the kink, at a value well below that of the
non-turbulent case at the end of the simulation (by a factor
of $\sim 2.4$). Unlike the non-turbulent case discussed in
the last paragraph, this leveling off cannot be due to magnetic
flux redistribution through the expansion of DEMS, which is
apparently absent in Model F (see Fig.\ \ref{DEMS}). For this model,
the magnetic flux levels off after the kink for the smaller
($\varpi \approx 3\times 10^{15}\cm$) circle as well, at a value lower
than that of the non-turbulent case by an even larger factor
($\sim 5$). The leveling off in the increase of magnetic flux is
a key to understanding the weakening of the DEMS by turbulence
and the appearance of the RSD\@. Since it starts around the kink
when the (perturbed) pseudodisk expands across a circle, it is
likely related to the structure of the pseudodisk, a possibility
that we will explore next.

\subsection{Turbulence-Warped Pseudodisk}
\label{Corrugation}

\begin{figure}
\epsscale{1.0}
\plotone{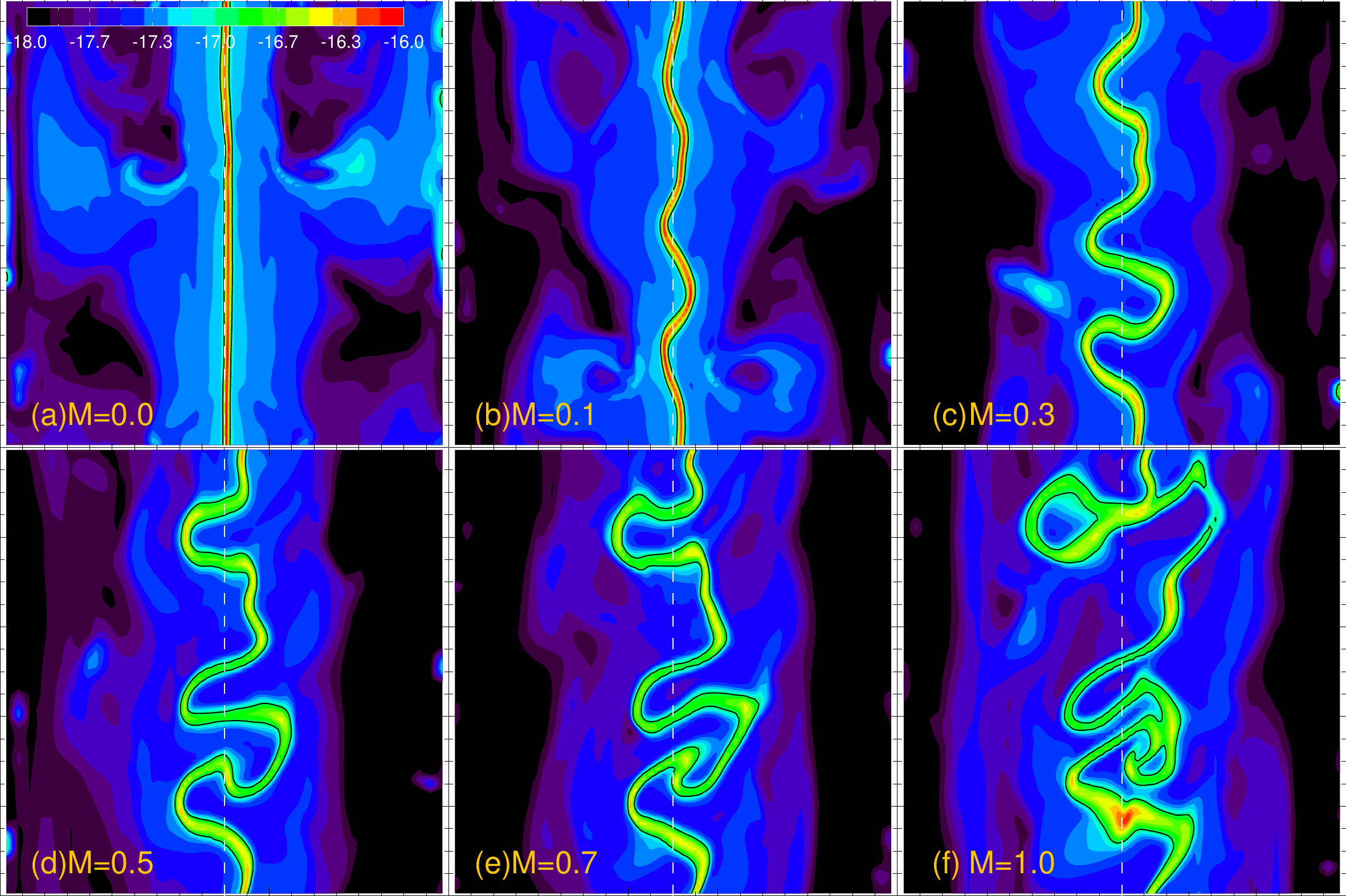}
\caption{Density distribution on a sphere of representative radius
$r=4.756\times 10^{15}\cm$ (or about $318\au$) at time $t=5\times
10^{11}\second$ for Models A--F\@. The horizontal axis is the polar angle
$\theta$ from $0$ to $\pi$, and vertical axis the azimuthal angle
$\phi$ from $0$ to $2\pi$. The equator plane at $\theta=\pi/2$ is
marked by a dashed line. The dense, equatorial, pseudodisk in the
non-turbulent model becomes increasingly distorted as the level of
turbulence increases.
The color map shows the logarithm of the density,
added to a floor value of $10^{-18}\rhounit$.
The black solid lines are constant density
contours at $10^{-17}\rhounit$ used to highlight the pseudodisk.
}
\label{DenSphere}
\end{figure}

In the absence of any turbulence, protostellar accretion in a dynamically
significant magnetic field is known to be controlled to a large extent
by the pseudodisk (\ct{GalliShu1993}). To highlight the morphology
of the pseudodisk and how it is perturbed by turbulence, we plot in
Fig.\ \ref{DenSphere} the density distribution on the surface of a
sphere at a representative radius of $r=4.756\times 10^{15}\cm$ for
Models A--F\@. It is evident that the mass in the
non-turbulent case is highly concentrated near the equatorial plane
($\theta=\pi/2$), in the pseudodisk. The mass concentration is
due to matter settling gravitationally along field lines toward the
equatorial plane, amplified by the compression by a severely pinched
magnetic field (for a pictorial view of the pseudodisk and associated
magnetic field in the non-turbulent case, see Fig.\ \ref{Pseudo}
below). This pseudodisk is
dynamically important because it
is where the majority of the core mass accretion occurs. For example,
in the non-turbulent case (Model A),
if we somewhat arbitrarily assign the region denser than
$10^{-17}\rhounit$ (bounded by the black solid lines in the
figure) to the pseudodisk,
then $85\%$ of the mass accretion at the
radius shown in Fig.\ \ref{DenSphere} occurs
through the pseudodisk, even though it covers only $2.7\%$ of the surface
area of the sphere. The concentration of mass accretion in the
pseudodisk in the non-turbulent case is an unavoidable consequence
of the interplay between the gravity and a dynamically significant,
ordered, magnetic field, whose opposition to the gravity is highly
anisotropic (\ct{GalliShu1993}; \ct{Allen+2003}).

The presence of a moderate level of turbulence does not change the
above picture fundamentally. For example, the $M=0.1$ turbulence in
Model B warps the nearly flat pseudodisk in the non-turbulent case
only slightly, as seen in panel (b) of Fig.\ \ref{DenSphere}.
As the level of turbulence
increases, the amplitude of pseudodisk warping grows. Nevertheless,
the pseudodisk retains its basic integrity even in the strongest
turbulence case of $M=1$ (Model F); it is severely distorted, with
some portions folding onto themselves (see panel f of the figure),
but not completely destroyed.

\begin{figure}
\epsscale{1.2}
\plottwo{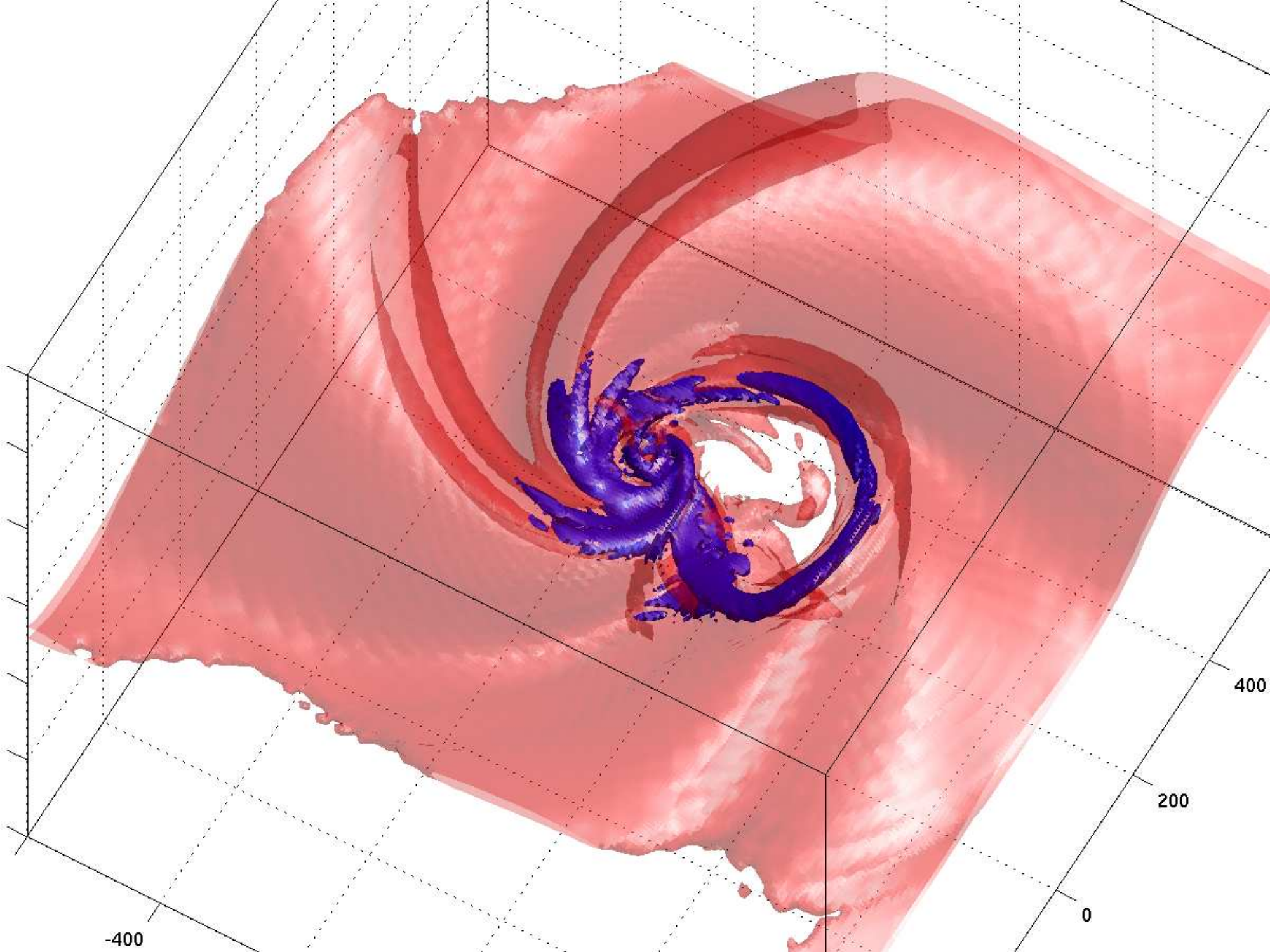}{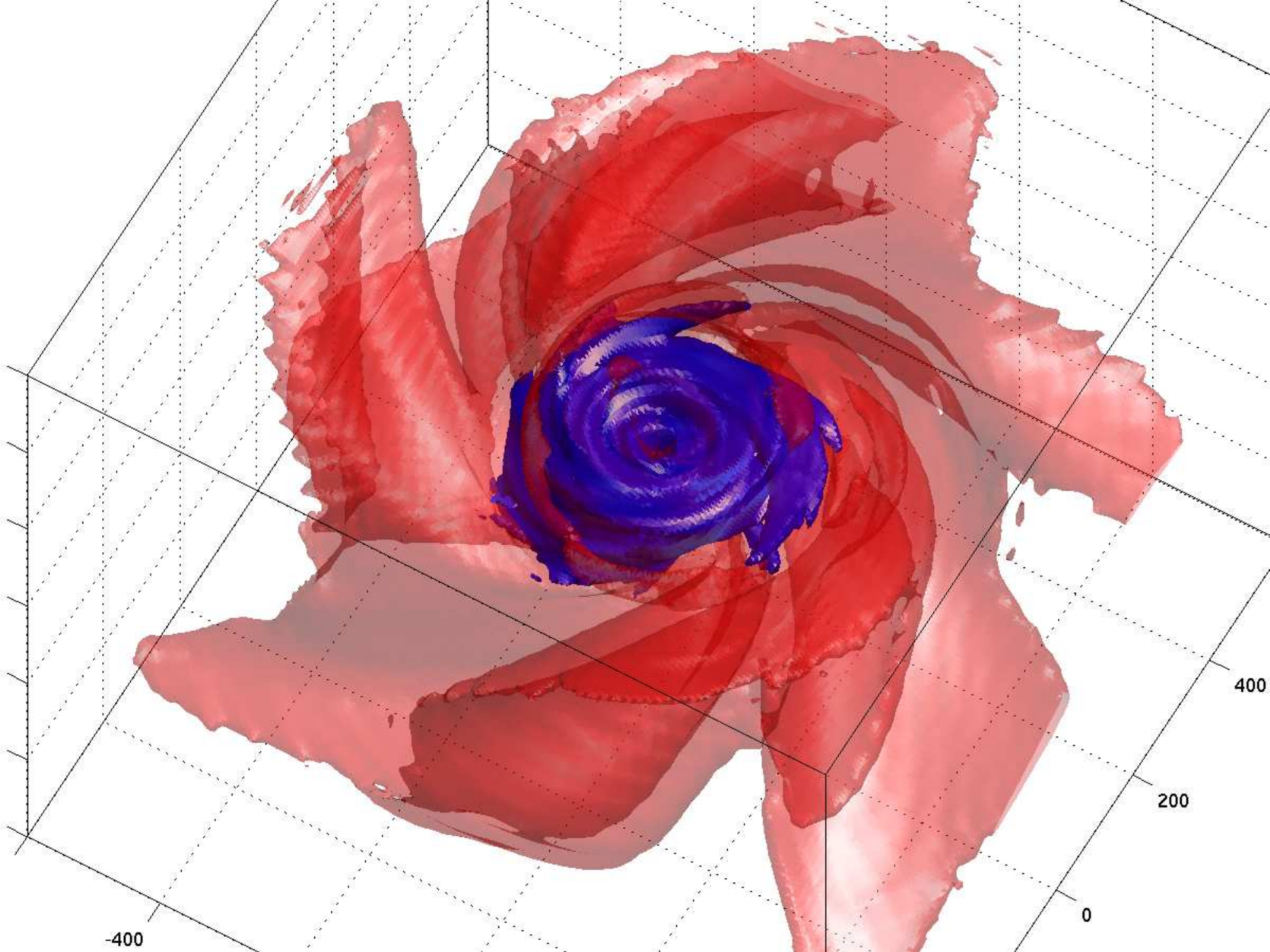}
\caption{3D structure of the warped pseudodisk. Plotted are isodensity
surfaces at $\rho=10^{-17}$ (red surfaces) and
$10^{-16}\rhounit$ (blue) for the $M=0.3$ (Model C, left panel) and $1$
(Model F, right panel) case at time $t=8\times 10^{11}\second$ (same as in
Figs.\ \ref{TurbE_6p} and \ref{DEMS}). Note the corrugation of the
pseudodisk (as traced by
the red surfaces) induced by turbulence. The blue region roughly
corresponds to the rotationally supported disk in the right panel
(Model F)\@. The low-density ``hole'' in the left panel
(Model C) corresponds to the strongly magnetized DEMS shown in the
upper right panel of Fig.\ \ref{DEMS}. The box size is $1200\au$ on each side.
}
\label{3DView}
\end{figure}

The turbulence-induced distortion of the pseudodisk can be viewed
more vividly in Fig.\ \ref{3DView}, where we plot two isodensity
surfaces at $\rho=10^{-17}$ and $10^{-16}\rhounit$ in 3D for
the $M=0.3$ and $1$ cases. At the time shown,
the corrugation induced by the subsonic, $M=0.3$
turbulence remains relatively moderate.
When the turbulent Mach number increases to 1, the pseudodisk, as
traced by the red isodensity surfaces, becomes more severely warped
and partially folded onto itself, but remains relatively thin.
In our simulations, the chaotic turbulent motion is dominated by the
fast, ordered, supersonic gravitational collapse in the region where the
pseudodisk is formed.
This, we believe, is the reason why the pseudodisk is perturbed,
rather than completely destroyed, by a subsonic or transonic
turbulence. In \S\ \ref{origin}, we will present general arguments for
the pseudodisk as a generic feature of magnetized core collapse.

The warped pseudodisk plays the same fundamental role in the
turbulent cases as the flat pseudodisk in the non-turbulent case: it
is the main conduit for core mass accretion. For example, on the spherical
surface
shown in Fig.\ \ref{DenSphere}, the warped pseudodisk (bounded by the
black density contours) is responsible for $72\%$, $69\%$, $74\%$,
$64\%$, and $68\%$ of the mass flux for the $M=0.1$, $0.3$, $0.5$, $0.7$,
and $1$ case, respectively, even though it covers only $3.3\%$,
$5.9\%$, $7.1\%$, $8.2\%$, and $12.2\%$ of the surface area. Since the
collapsing pseudodisk is mainly responsible for the mass accretion
that drags the field lines into the circumstellar region close to the central
object, it should not be too surprising that its distortion by
turbulence affects the magnetic flux accumulation there, as we show
next.

\subsection{Pseudodisk Warping and Magnetic Flux Reduction}
\label{FluxLoss}

As discussed in \S\ \ref{MagFlux} and illustrated in Fig.\ \ref{Bflux},
turbulence tends to lower the magnetic flux
threading the equatorial plane at small radii at late times. To understand this trend quantitatively, we note
that the evolution of the magnetic flux $\Phi_z$ enclosed within a
circle of fixed cylindrical radius $\varpi$ on the equatorial plane
is governed by the
induction equation,
which can be cast into the following form using the Stokes theorem
\begin{equation}
{\partial\Phi_z\over\partial t} = \int_0^{2\pi} E_\phi\ \varpi\ d\phi
  = - \int_0^{2\pi} v_\varpi B_z\ \varpi\ d\phi + \int_0^{2\pi} v_z
  B_\varpi\ \varpi\ d\phi,
\label{induction}
\end{equation}
in a cylindrical coordinate system $(\varpi, \phi, z)$. The quantity
$E_\phi = -v_\varpi B_z + v_z B_\varpi$ is the azimuthal component of
the electromotive force (EMF) on the circle. On the equatorial plane,
the relevant components of the velocity and magnetic field in
cylindrical and spherical coordinates are related through
$v_\varpi=v_r$, $B_\varpi=B_r$, $v_z=-v_\theta$ and $B_z=-B_\theta$.
The first term on the right hand side of the above equation,
$T_{\phi,r}=\int_0^{2\pi} v_r B_\theta\ \varpi\ d\phi$, has an
obvious interpretation: it is simply the rate of flux
advection by radial infall, which tends to increase the
flux $\Phi_z$ (and thus be positive) by dragging vertical
field lines into the circle. The meaning of the second term
$T_{\phi,z}= - \int_0^{2\pi}v_\theta B_r\ \varpi\ d\phi$ is less
obvious; it is the rate of flux advection by vertical motions
(along the $z$-axis, perpendicular to the equatorial plane)
that can move radial field lines across the circle on the
equatorial plane.

\begin{figure}
\epsscale{1.0}
\plotone{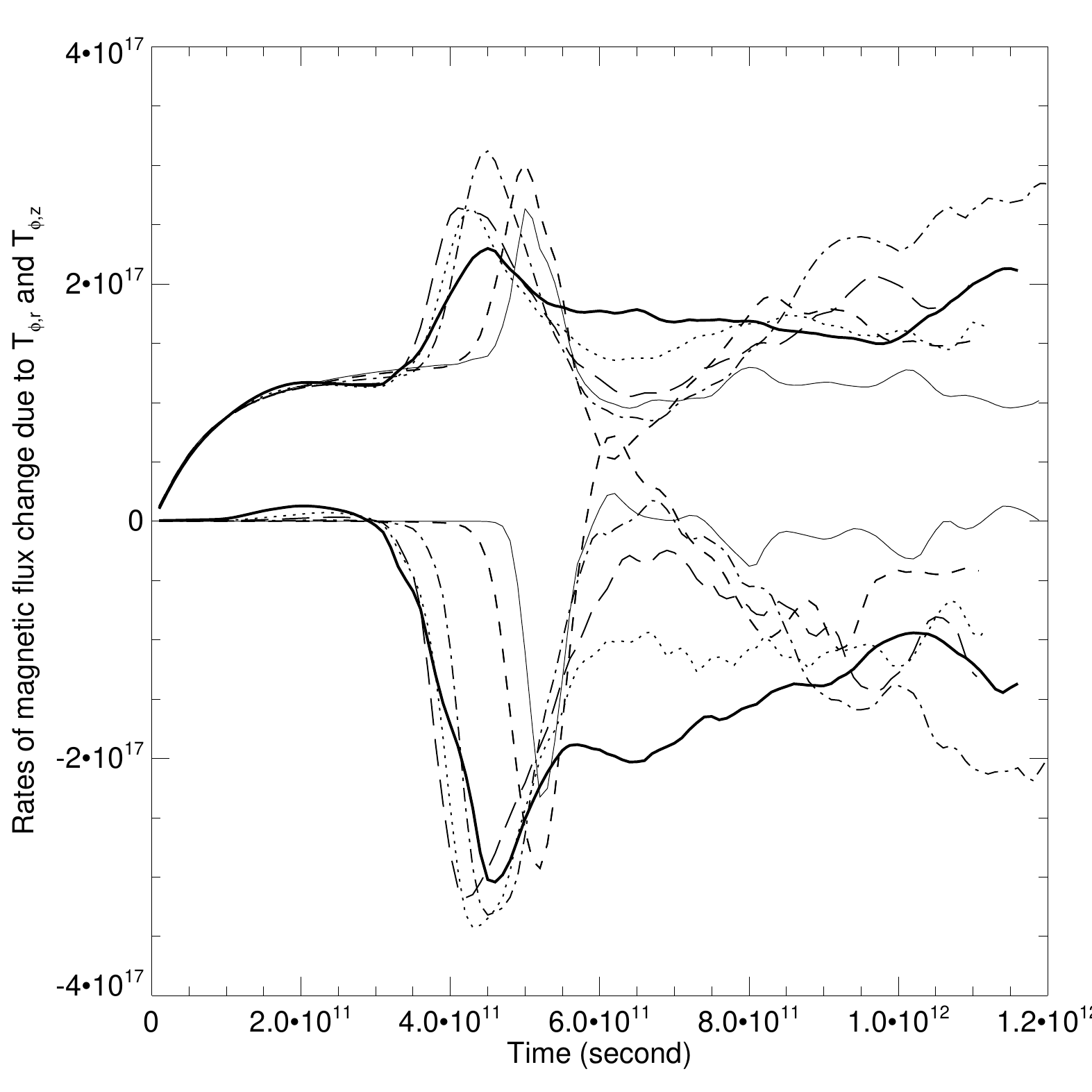}
\caption{The rates of magnetic flux change, $T_{\phi,r}$ and
$T_{\phi,z}$ (in cgs Gaussian units), across a circle of radius
$\varpi=1.055 \times 10^{16}\cm$
on the equatorial plane due to, respectively, radial advection
of vertical field $B_\theta$ by infall (with $v_r$; upper curves)
and vertical advection of radial field $B_r$ by vertical motions
(with $v_\theta$; lower curves), for Model A ($M=0$, thin
solid line), B (dashed), C (dash-dotted), D (long dashed), E (dotted)
and F ($M=1$, thick solid line). Note that turbulence increases the rate of
outward (or negative) magnetic flux advection by vertical motions.
}
\label{EMF}
\end{figure}

It is easy to compute $T_{\phi,r}$ and $T_{\phi,z}$ for any radius
$\varpi$. As an example, we plot in Fig.\ \ref{EMF} their values at
$\varpi=1.055 \times 10^{16}\cm$ as a function of time for different models.
At early times, the radial flux advection term $T_{\phi,r}$
dominates the vertical flux advection term $T_{\phi,z}$ for
all cases. This is to be expected, because the field lines
near the equator remain predominantly vertical outside the
pseudodisk (see Figs.\ \ref{WarpB} and \ref{Pseudo}
below). After the outer edge of the pseudodisk passes through
the circle, the magnetic
fluxes in the non-turbulent and weakly turbulent cases start
to increase again (see the left panel of Fig.\ \ref{Bflux}).
This is because the vertical flux advection that tends to move field
lines out of the circle (i.e., $T_{\phi,z}$ tends to be
negative\footnote{This is
because in the pseudodisk region a highly pinched
field configuration tends to develop, with a generally positive $B_r$
above the equatorial plane and negative $B_r$ below it (see
Figs.\ \ref{Pseudo} and \ref{WarpB} for illustration). A
downward motion (with a positive $v_\theta$) tends to push the
field lines in the upper hemisphere (which generally point
radially outward, with a positive $B_r$) downward across the circle on
the equatorial plane, and an upward motion (with a negative
$v_\theta$) tends to push the field lines in the lower
hemisphere (which generally point radially inward, with a
negative $B_r$) upward across the circle. In both cases,
the product $-v_\theta B_r$ tends to be negative, indicating
that vertical motions tend to move magnetic flux out of the
circle.}), starts to drop below the radial flux advection that
tends to move field lines into the circle. An exception
is the strongest turbulence case of $M=1$, where the vertical
and radial advection terms stay comparable, so that the magnetic
flux changes relatively little at late times. Fig.\ \ref{EMF}
shows clearly that the more efficient
outward transport of magnetic flux by vertical motions is the main
reason for the slower flux increase for a stronger turbulence.
In order for the outward flux transport to be efficient, both
$v_\theta$ and $B_r$ need to have a relatively large value,
which can be achieved when a pseudodisk with a highly
pinched magnetic field (i.e., an appreciable $B_r$) is strongly
perturbed vertically (i.e., an appreciable $v_\theta$), by
turbulence or some other means.

To illustrate how a strongly perturbed pseudodisk can slow down the
magnetic flux accumulation inside a circle more pictorially, we plot
in Fig.\ \ref{WarpB} the density distribution and magnetic field
(unit) vectors on a representative meridian plane for the $M=1$
case. Note that the
turbulence-driven warping moves the pseudodisk above (see the loop
to the right of the disk) and below (see the loop to the left) the
equatorial plane. The above-the-equatorial-plane loop (on the right
side) delivers a
substantial amount of matter through the upper hemisphere (marked
by the dashed line in the figure) but little net magnetic flux; the
sharp kink of field lines across the loop allows most of the field
lines dragged into the hemisphere by the accreting loop to
return through the same hemisphere, without crossing or touching the
circle on the equatorial plane (marked by two red crosses on the
figure); a similar case
can be made for the lower hemisphere. This is in contrast with the
mass accretion through the unperturbed equatorial pseudodisk, which
must be accompanied by a flux increase (in the ideal MHD limit).
Since the net magnetic flux
going through the upper (or lower) hemisphere is the same as that
through the equatorial plane that bisects the sphere, the pseudodisk
warping provides a natural explanation for the lower magnetic flux
accumulated close to the central object on the equatorial plane
(and thus a weaker DEMS) that we found for a stronger turbulence.

\begin{figure}
\epsscale{1.0}
\plotone{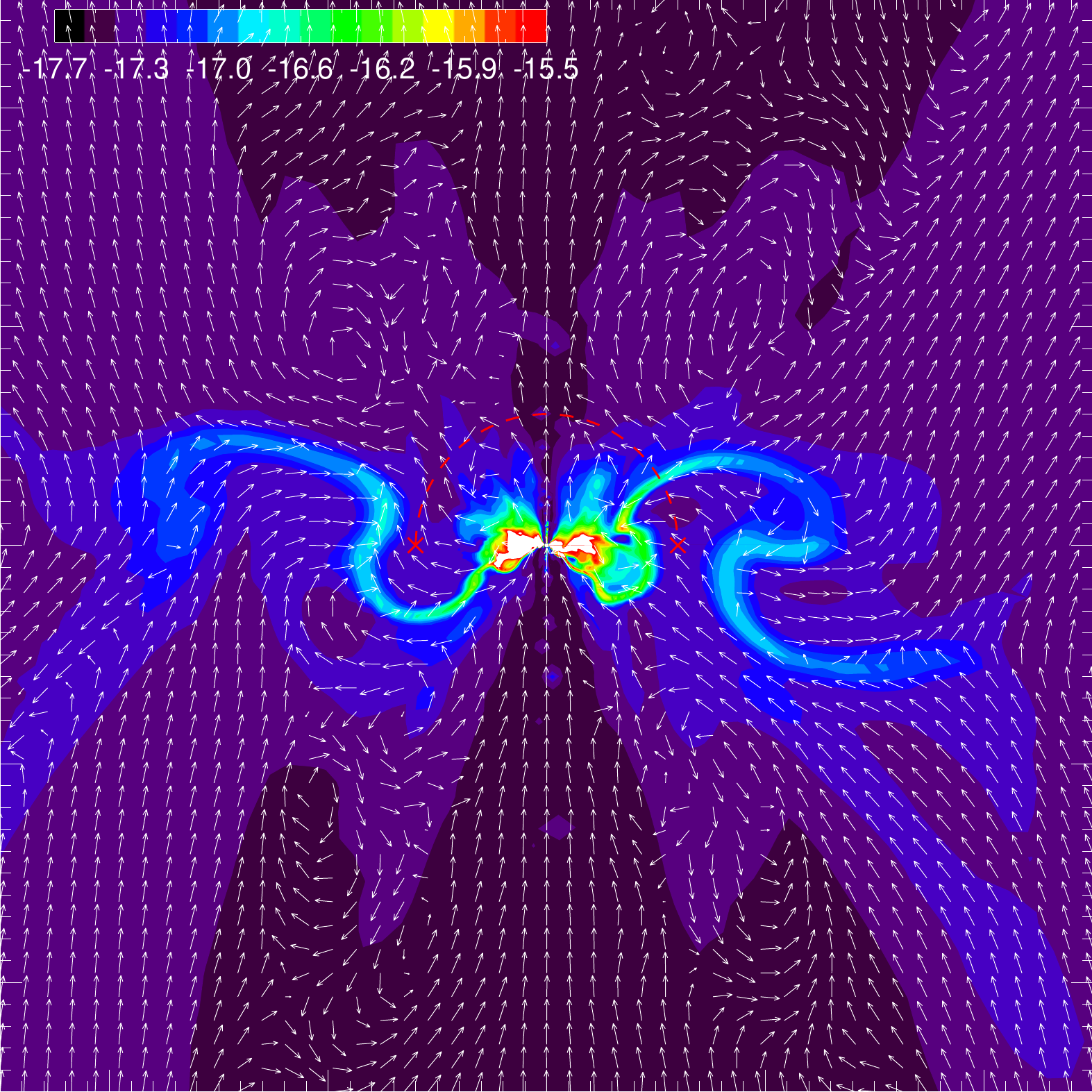}
\caption{Density map and magnetic field unit vectors of the $M=1$
model on a meridian plane at a representative time
$t=8\times 10^{11}\second$. It illustrates how an
out-of-the-equatorial-plane dense loop (the loop on the
right side, part of the warped pseudodisk)
can bring matter through the upper hemisphere (dashed line) but little
magnetic flux. The two crosses mark where the hemisphere and
equatorial plane intersect. The length of the box is $5\times
10^{16}\cm$ on each side. The colorbar is as in Fig.\ \ref{TurbE_6p}.
}
\label{WarpB}
\end{figure}

\subsection{Torque Analysis}
\label{Torque}

Whether a rotationally supported disk can form or not depends on the
amount of angular momentum that is initially available on the core
scale and that is removed by magnetic torque and outflow as the rotating core
material collapses toward the central object. We follow
\citet{Li+2013} in evaluating the $z$-components of the dominant magnetic
torque due to the magnetic tension force (as opposed to the magnetic
pressure gradient) and the advective torque:
\begin{equation}
N_{t,z} = \frac{1}{4\pi} \int \varpi B_\phi B_r\ dS,
\end{equation}
and
\begin{equation}
N_{a,z} = -\int \rho \varpi v_\phi v_r\ dS,
\end{equation}
where $\varpi$ is the cylindrical radius, and the integration is
over the surface $S$ of a sphere of radius $r$. They measure,
respectively, the rate of angular momentum change in the volume
enclosed by the surface $S$ due to magnetic braking and matter
crossing the surface $S$. The advective torque consists of two parts: the
rates of angular momentum advected into and out of the sphere
by infall and outflow respectively:
\begin{equation}
N_{a,z}^{\rm{in}} = -\int \rho \varpi v_\phi v_r (< 0)\ dS,
\end{equation}
and
\begin{equation}
N_{a,z}^{\rm{out}} = -\int \rho \varpi v_\phi v_r (> 0)\ dS.
\end{equation}

\begin{figure}
\epsscale{1.25}
\plottwo{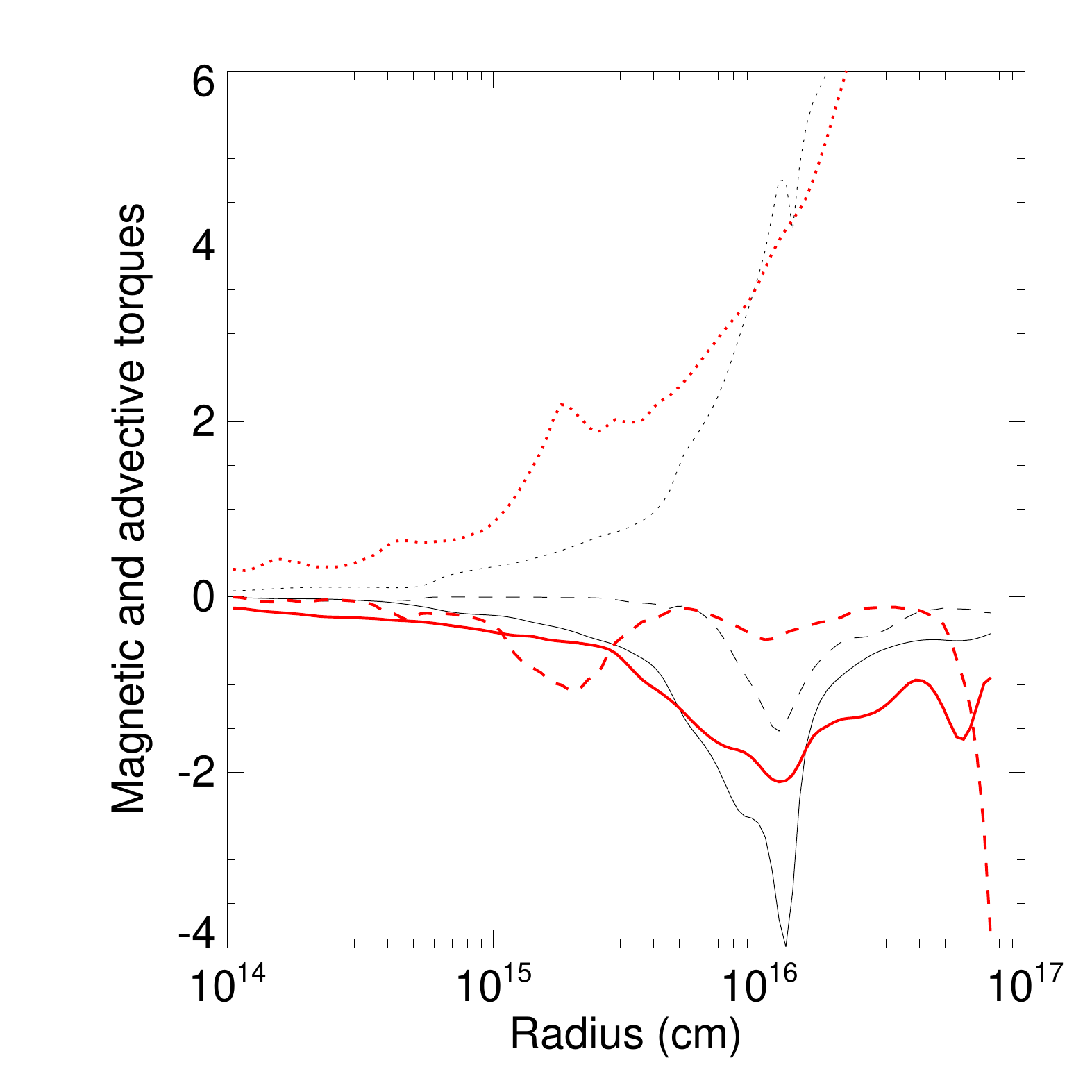}{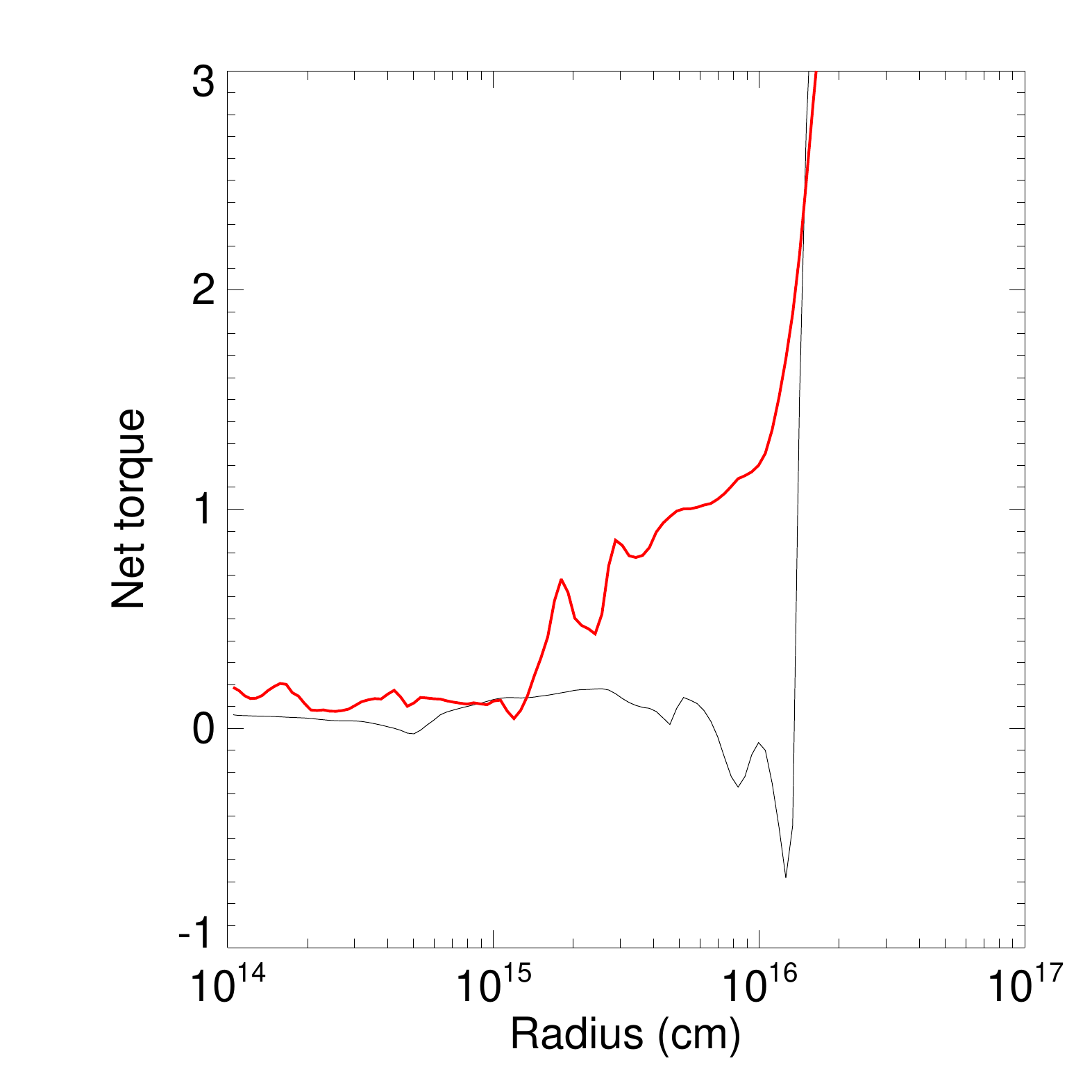}
\caption{Left panel: Magnetic torques (solid lines, $N_{t,z}$) and advective
torques ($N_{a,z}$) by infall ($N_{a,z}^{\rm{in}}$, dotted) and
outflow ($N_{a,z}^{\rm{out}}$, dashed) acting
on spheres of different radii for models with $M=0$ (thin black lines)
and 1 (thick red lines), at a representative time $t=7\times
10^{11}\second$ for Model F\@.
Right panel: The net torque ($N_{t,z}+N_{a,z}$) for the same two cases.
The torques are in units of $10^{40}\dyn\cm$.
}
\label{Torq}
\end{figure}

We have examined the radial distributions of the magnetic and
advective torques for Models A--F and at different
times.
The basic features of the distributions are well illustrated in
the examples shown in Fig.\ \ref{Torq}. To avoid crowding,
we have shown only two extreme cases (with $M=0$ and $1$).
The right panel of the figure, which displays the net torque,
shows that, in the non-turbulent
case, the magnetic torque is large enough to remove essentially
all of the angular momentum advected inward at all
radii inside $\sim 10^{16}\cm$; indeed, it is so strong as to
cause a net decrease in angular momentum between $\sim 3\times
10^{15}$ and $\sim 10^{16}\cm$. This latter feature is in
striking contrast with the $M=1$ case,
where the magnetic torque is not large enough to remove all of
the angular momentum brought in by flows. The imbalance
leaves a substantial net (positive) torque between $\sim
3\times 10^{15}$ and $\sim 10^{16}\cm$, which increases the
angular momentum of the material in this region, enabling
a rotationally supported disk to form in this case. Since
the difference between the two cases appears most prominent
near $r \sim 10^{16}\cm$, we will first focus on this region in
our effort to understand
why the magnetic braking is so efficient in the non-turbulent
case and why the efficiency is significantly decreased by the
$M=1$ turbulence.

\begin{figure}
\epsscale{1.0}
\plotone{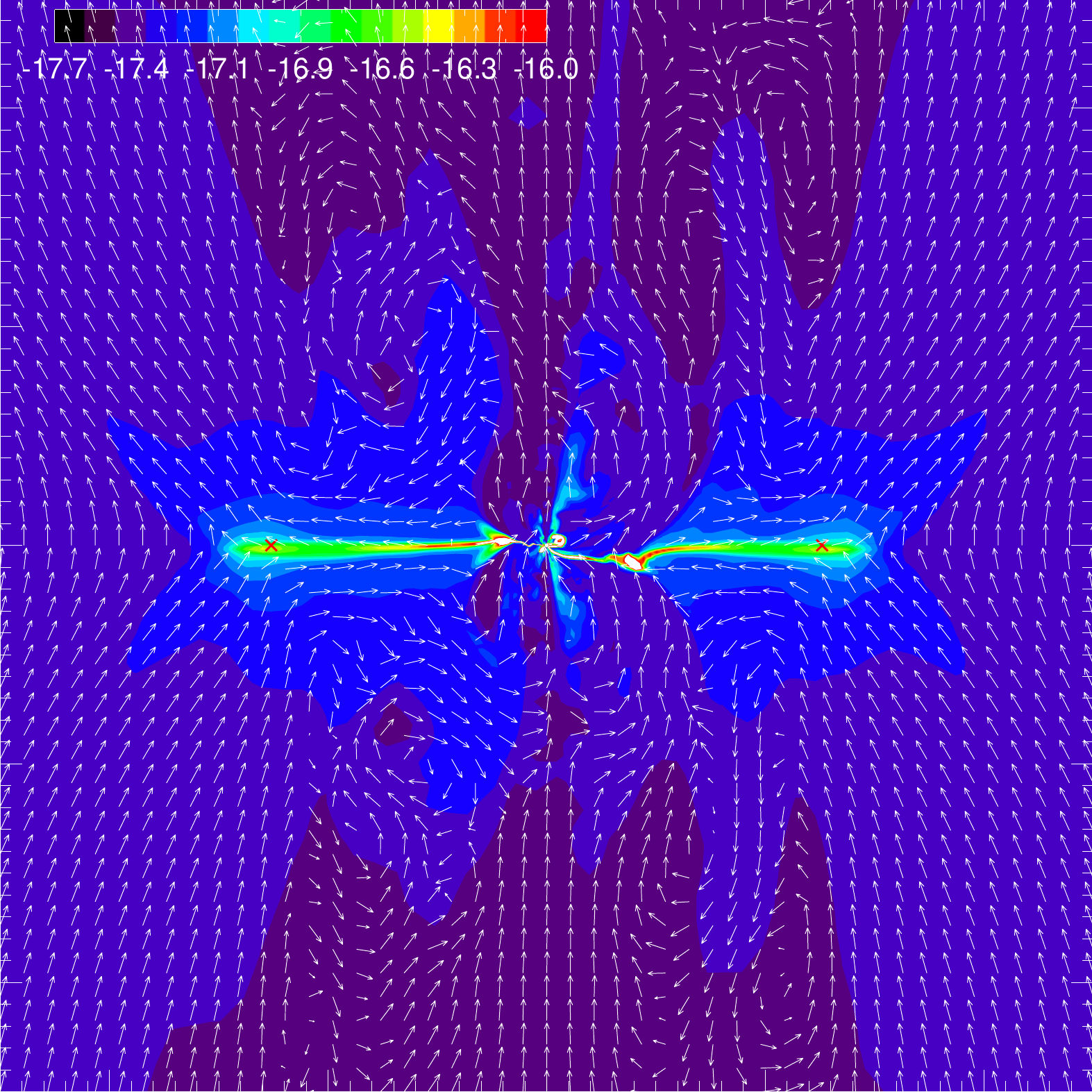}
\caption{Density map and magnetic field unit vectors on a
meridian plane for the non-turbulent case at a representative time
$t=7\times 10^{11}\second$, showing a prominent equatorial pseudodisk and
severe field pinching across it (an axial dense spot has been erased for better clarity of the equatorial region).
The two crosses mark the locations
where the magnetic torque shown in Fig.\ \ref{Torq} peaks.
The colorbar is as in Fig.\ \ref{TurbE_6p}.
The length of the box is $5\times
10^{16}\cm$ on each side.
}
\label{Pseudo}
\end{figure}

\begin{figure}
\epsscale{1.25}
\plottwo{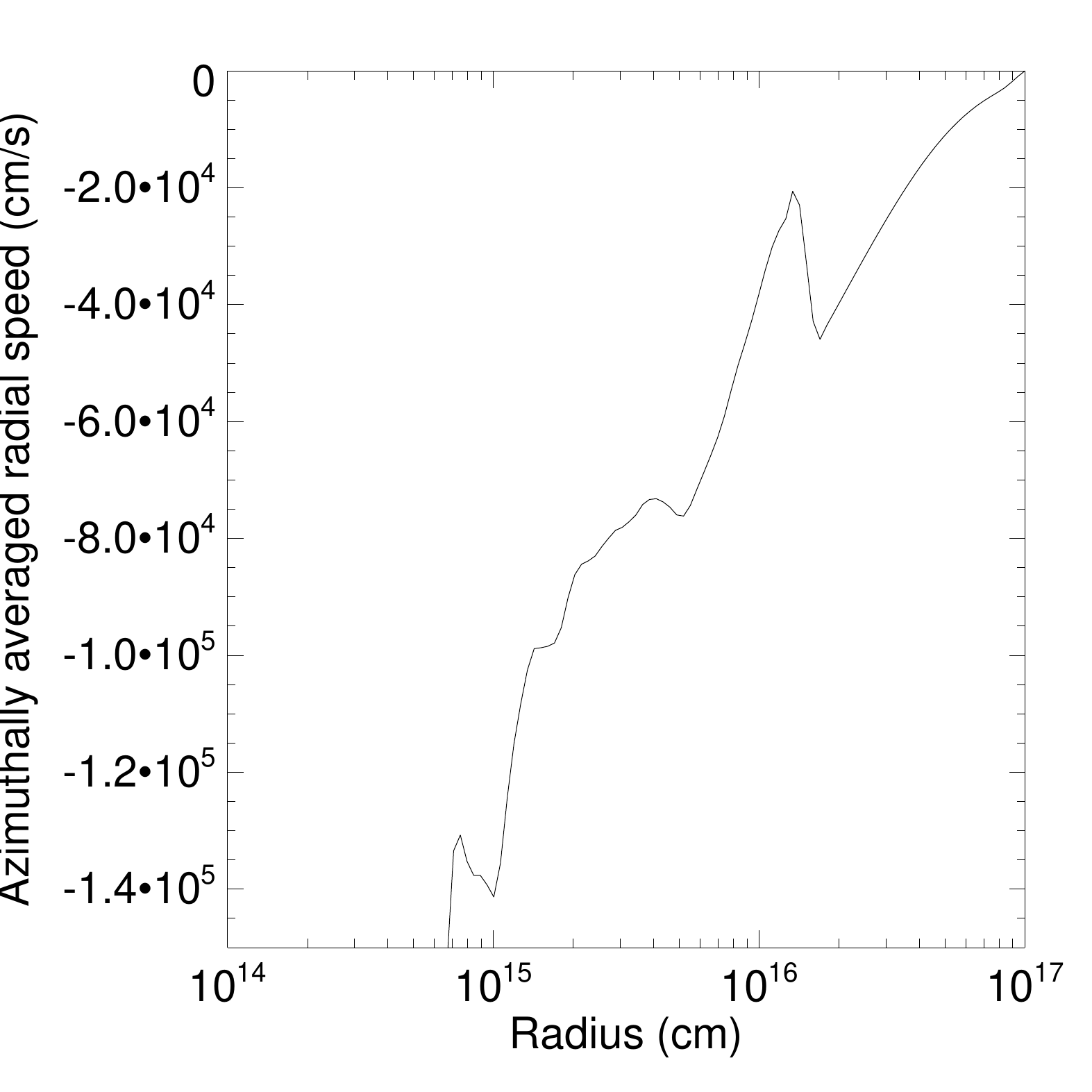}{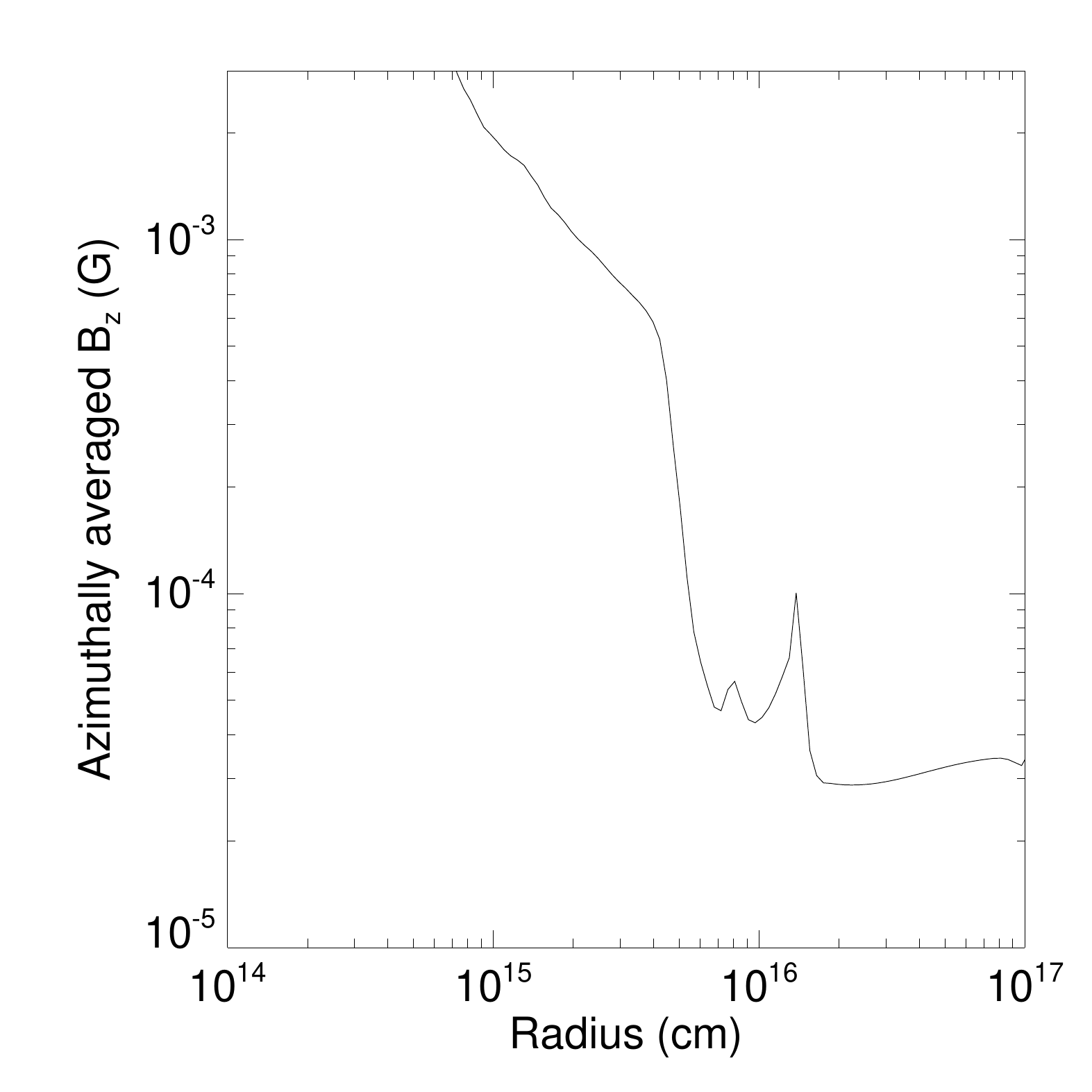}
\caption{Distributions of the azimuthally averaged infall speed
($v_r$, left panel) and vertical field strength ($B_z$, right)
on the equatorial plane for
the non-turbulent case shown in Fig.\ \ref{Pseudo}. Note the sharp
slow-down of infall and increase in field strength just outside
$r=10^{16}\cm$, precisely where the magnetic torque peaks.
}
\label{Barrier}
\end{figure}

We first concentrate on the non-turbulent case. It turns out that
the $r\sim 10^{16}\cm$ region is rather special; it includes the
outer part of the pseudodisk. This is illustrated in
Fig.\ \ref{Pseudo}, where we display the density map on a meridian
plane (which shows the pseudodisk clearly) and unit vectors for
the magnetic field at the same representative time as in
Fig.\ \ref{Torq}. The two crosses mark the locations where the
magnetic torque peaks (at $r\approx 1.3\times 10^{16}\cm$).
It is clear that, as matter enters the
equatorial pseudodisk, it drags the field lines into a highly pinched
configuration (note the reversal of the radial field above and
below the pseudodisk). Associated with the pinch is a large magnetic
tension force in the radial direction, which acts against the gravity
and retards the collapse significantly. The retardation can be seen in
Fig.\ \ref{Barrier}, which shows that the azimuthally averaged infall
speed on the equator is suddenly reduced by about a factor of two
right outside $r\sim 10^{16}\cm$, precisely where the rate of magnetic
braking peaks. This region of sharp deceleration of the magnetized
collapsing flow is termed the ``magnetic barrier'' by
\citeauthor{MellonLi2008} (\citeyear{MellonLi2008}, see their Fig.\ 4);
this barrier is analogous to the well-known ``centrifugal barrier''
where the infall is quickly slowed down by rotation. The slow-down
allows both matter and magnetic field lines to pile up, signaling
the formation of a dense, strongly magnetized pseudodisk. The pileup
of field lines can be seen in the right panel of Fig.\ \ref{Barrier},
which shows that the vertical component of the magnetic field on
the equator, $B_z$, increases sharply by a factor of $\sim 3$ at
the magnetic barrier.\footnote{Inside the barrier, $B_z$ drops somewhat as
the material inside the pseudodisk re-accelerates inward. The region
of strong magnetic field inside a radius of $\sim 4\times 10^{15}\cm$
corresponds to the DEMS that is visible in Figs.\ \ref{TurbE_6p}
and \ref{DEMS}.} The increased field
strength, coupled with severe field pinching (which increases the
lever arm for magnetic braking, see Fig.\ \ref{Pseudo}),
is the reason behind the efficient braking at the magnetic barrier
in the non-turbulent case.

In the presence of a sonic turbulence ($M=1$), the peak rate of
angular momentum removal by magnetic torque is substantially
reduced (by a factor of $\sim 2$, see Fig.\ \ref{Torq}). Our
interpretation is that the
reduction is due to the distortion of the highly coherent
magnetic barrier of the $M=0$ case by turbulence. Fig.\ \ref{WarpB}
shows that the transition from the infall envelope to the
pseudodisk is less coherent and more gradual in the $M=1$ case
compared to the $M=0$ case (Fig.\ \ref{Pseudo}). As a result,
the rotation is braked more gently as the matter enters the
(highly warped) pseudodisk. The weaker
braking at the outer part of the pseudodisk leaves the material
inside the pseudodisk with more angular momentum, making it more
likely to form a rotationally supported disk.

Another difference between the $M=0$ and $1$ case lies in the
outflow. In the non-turbulent case, the outflow is driven mostly
by the equatorial (rotating) pseudodisk, which winds up the
field lines, building up a magnetic pressure near the equatorial
plane that is released by (bipolar) expansion away from the
plane. On the scale of the pseudodisk ($\sim
10^{16}\cm$) that is crucial for disk formation, the outflow
removes angular momentum at a rate that is a substantial fraction
(typically $\sim 1/3$ to $1/2$) of that by magnetic torque (see the
left panel of Fig.\ \ref{Torq} for an example). This is in contrast
with the
$M=1$ case, where the angular momentum removal by outflow is much
less efficient on the same scale (see Fig.\ \ref{Torq}). The
lower efficiency is most likely caused by the severe warping of
the pseudodisk, which weakens the ability of the rotating
material in the warped pseudodisk to generate a coherent toroidal
field for outflow driving (a similar point was also made in
\ct{Seifried+2012,Seifried+2013}). Furthermore,
the outflow, if driven at all,
will consist of strands coming from different parts of the warped
pseudodisk, which may have different orientations; strands moving
in different directions may lead to cancellation that weakens the
net efficiency of the outward angular momentum transport by the
outflow. We should note that, at smaller radii (on the $\sim
10^{15}\cm$, rotationally supported disk-scale), the outflow in
the $M=1$ case removes
angular momentum more efficiently than that in the $M=0$
case. This outflow is driven by the RSD, and is thus a consequence of, rather than the cause for,
the disk formation. Nevertheless, it removes angular momentum from
the RSD, and could threaten its survival; it may have contributed
to the destruction of the transient disks in the $M=0.5$ and $0.7$
cases (Models D and E in Table 1).

\subsection{Varying Initial Turbulent Velocity Field}

\begin{figure}
\epsscale{1.0}
\plotone{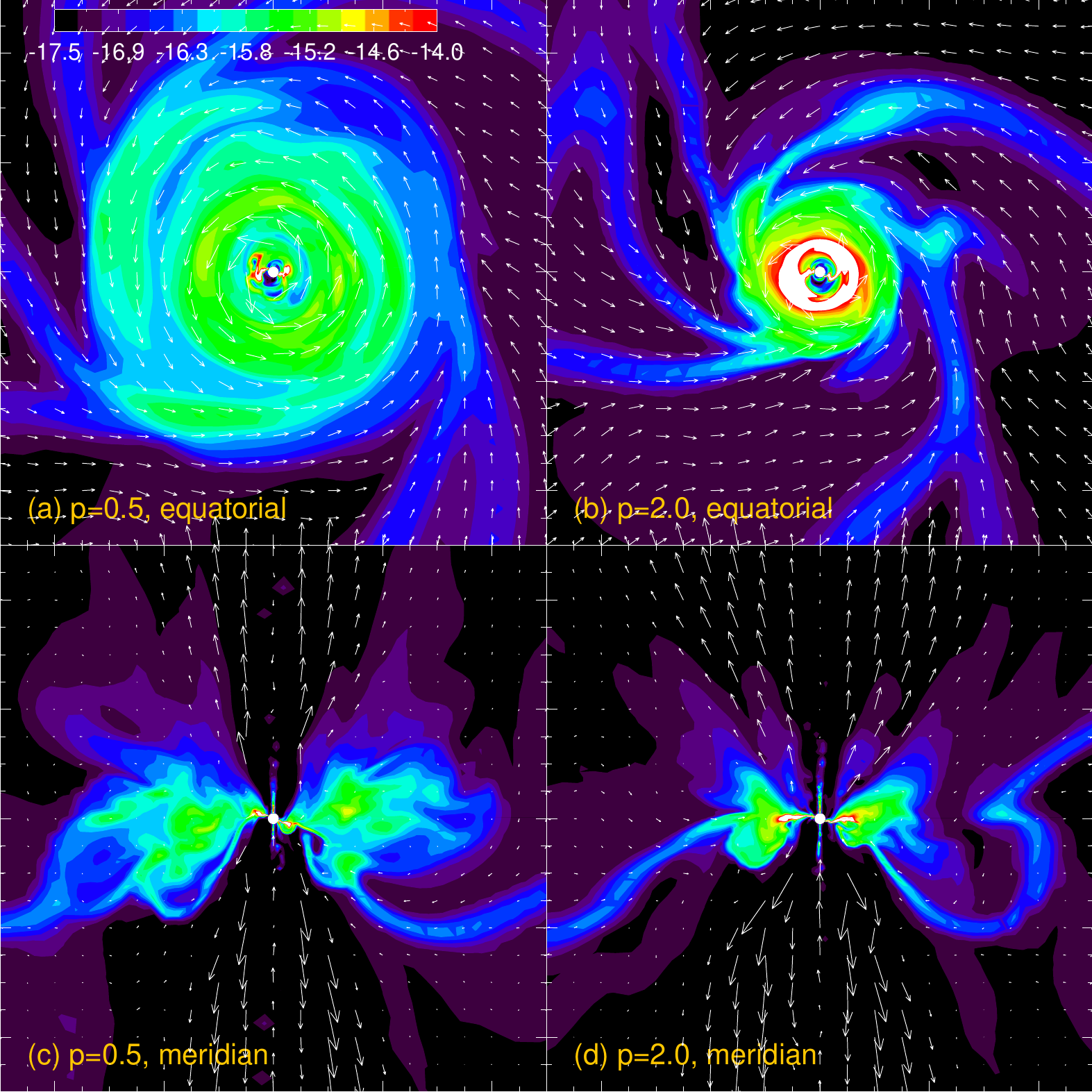}
\caption{Density map and velocity field for Model U ($p=0.5$, left panels) and V
($p=2.0$, right panels) on the equatorial (top panels) and a meridian (bottom
panels) plane at time $t=6\times 10^{11}\second$. A well developed disk
is apparent in both cases.
The colorbar is as in Fig.\ \ref{TurbE_6p}.
The length of the box is $10^{16}\cm$ on each side.
}
\label{power}
\end{figure}

We have carried out several shorter duration simulations with
different turbulent velocity fields to explore their effects on
disk formation. Two examples are shown in Fig.\ \ref{power}. They
are identical to Model F (with $M=1$ and an exponent for the
turbulent velocity spectrum $p=1$), except for $p=0.5$ (Model
U) or $2.0$ (Model V)\@. In both cases, a rotationally supported
disk is formed at the time shown ($t=6\times 10^{11}\second$),
just as in Model F\@. The disk is somewhat larger and better
developed
in Model U than in Model V, indicating that the shallower
turbulent velocity spectrum (with more power at shorter
wavelength) is more conducive to disk formation.
However, we refrain from drawing more quantitative
conclusions because the turbulent velocity fields are distorted
by our non-uniform grid due to the undersampling of the high-frequency
part of the velocity spectrum at large radii, where the spatial
resolution is relatively coarse.

\section{Discussion}
\label{discussion}

\subsection{Unification of Turbulence- and Misalignment-Enabled Disk
Formation}
\label{Unification}

\begin{figure}
\epsscale{1.0}
\plottwo{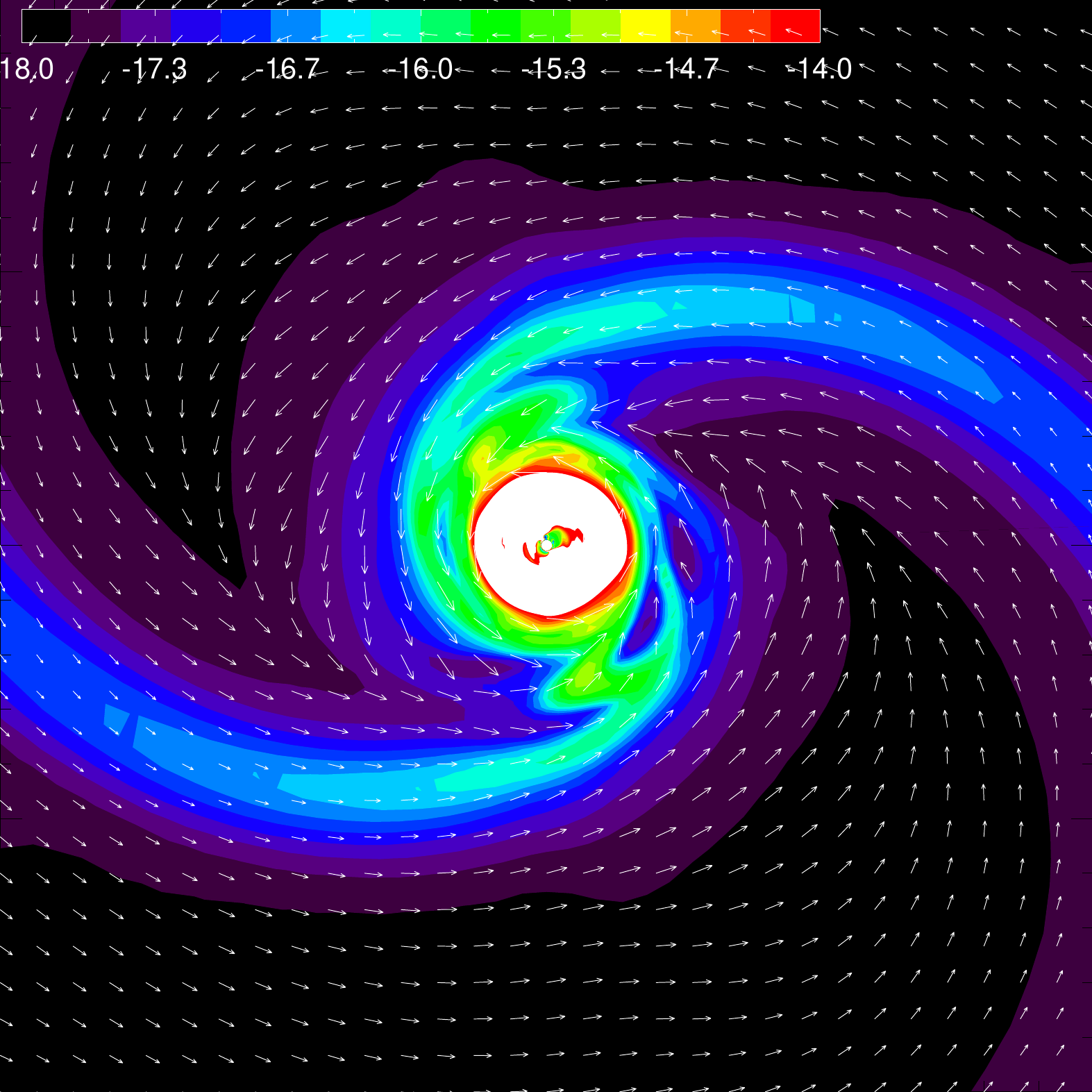}{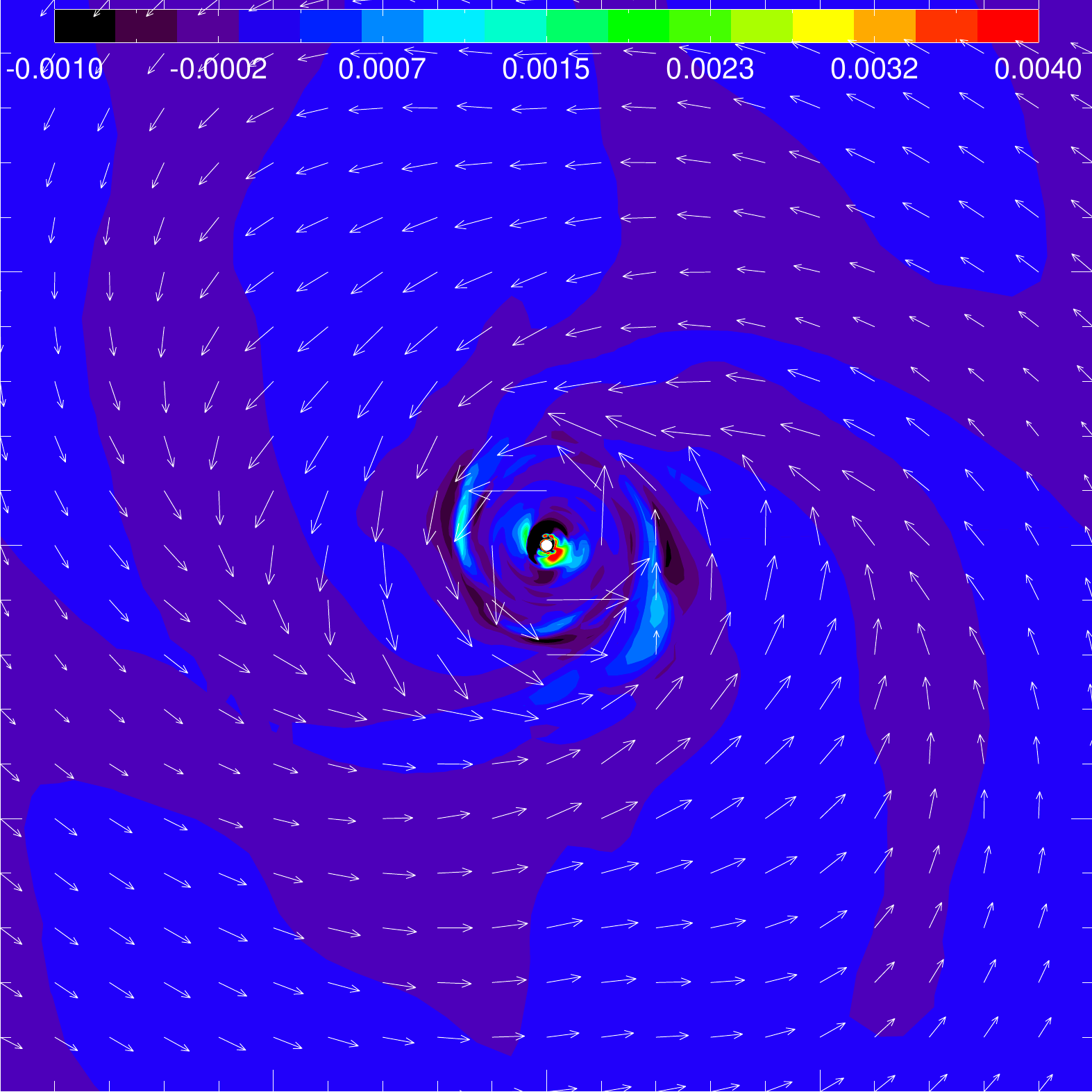}
\caption{Distribution of the logarithm of density (in $\rhounitnosp$) and
velocity field (left panel) and the vertical component of the magnetic
field $B_z$ and velocity field (right panel) on the equatorial
plane at a representative time $t=8\times 10^{11}\second$ for Model
P where the magnetic field is perpendicular to the rotation axis. The
two prominent spirals in the left panel are part of the strongly
warped pseudodisk that
feeds the central rotationally supported disk. The central, white
part has a density above the maximum value for the color plot,
which is set to a relatively low value in order to highlight the
spirals. The right panel shows that the strongly magnetized DEMS
that prevented disk formation in the aligned, non-turbulent and weakly
turbulent cases (see Fig.\ \ref{DEMS}) disappears almost completely,
strengthening the case for the elimination of DEMS as a prerequisite
for disk formation.
The length of the box is $2\times 10^{16}\cm$ on each side for both
panels.
}
\label{t90}
\end{figure}

The warping of pseudodisk out of the disk-forming equatorial plane by
turbulence plays a central role in our interpretation of the robust
disk formation observed in our simulations. Pseudodisk warping was
also the key ingredient of another proven mechanism for disk formation:
misalignment between the magnetic field and rotation
axis (\ct{HennebelleCiardi2009}; \ct{Joos+2012}; \ct{Krumholz+2013};
\ct{Li+2013}). Since our problem setup is somewhat different from those of
previous studies (they included self-gravity
that is ignored here, see \S\ \ref{setup}), we have rerun the non-turbulent
case (Model A) but with the magnetic field perpendicular to the
rotation axis (Model P in Table 1). A robust rotationally supported
disk is easily formed in this case, as shown in the left panel of
Fig.\ \ref{t90}.
As in the case with self-gravity,
there are two prominent spiral arms in the equatorial density map of
Fig.\ \ref{t90}, which are part of a pseudodisk that lies almost
perpendicular to the equatorial plane initially and is wrapped by
rotation into a snail shell-like structure\footnote{In this paper, we will
call the snail shell-like structure a pseudodisk even though it is
not disk-like, because it is produced by magnetically channeled
gravitational collapse, just as the unperturbed (flat) pseudodisk.}
in 3D (see Fig.\ 2 of \ct{Li+2013}).

Although the warping of the pseudodisk in Model P is more extreme and
less chaotic than that induced by the sonic turbulence ($M=1$) in
Models F, U and V, the underlying physical reason for disk formation
and survival appears broadly similar. Specifically, the large
field-rotation misalignment ensures that the bulk of the pseudodisk
material stays out of the equatorial plane, which alleviates the
problem of magnetic flux ($\Phi_z$) accumulation on the equatorial
plane. This in turn eliminates the highly magnetized DEMS that is
detrimental to disk formation (\ct{Zhao+2011};
\ct{Krasnopolsky+2012}), as shown in the right panel of
Fig.\ \ref{t90}.
In addition, the misalignment decreases the rate of angular momentum
removal from the pseudodisk by outflow (\ct{CiardiHennebelle2010};
\ct{Li+2013}) and weakens the braking near the magnetic barrier, both
of which leave more angular momentum in the accretion flow to form RSDs.
The turbulence-
and misalignment-enabled disk formation are thus unified, in that both
cause the pseudodisk to warp strongly out of the equatorial plane
(defined by rotation), which is conducive to disk formation.

\subsection{Origin of Rotationally Supported Disks}
\label{origin}

\begin{figure}
\epsscale{0.5}
\plotone{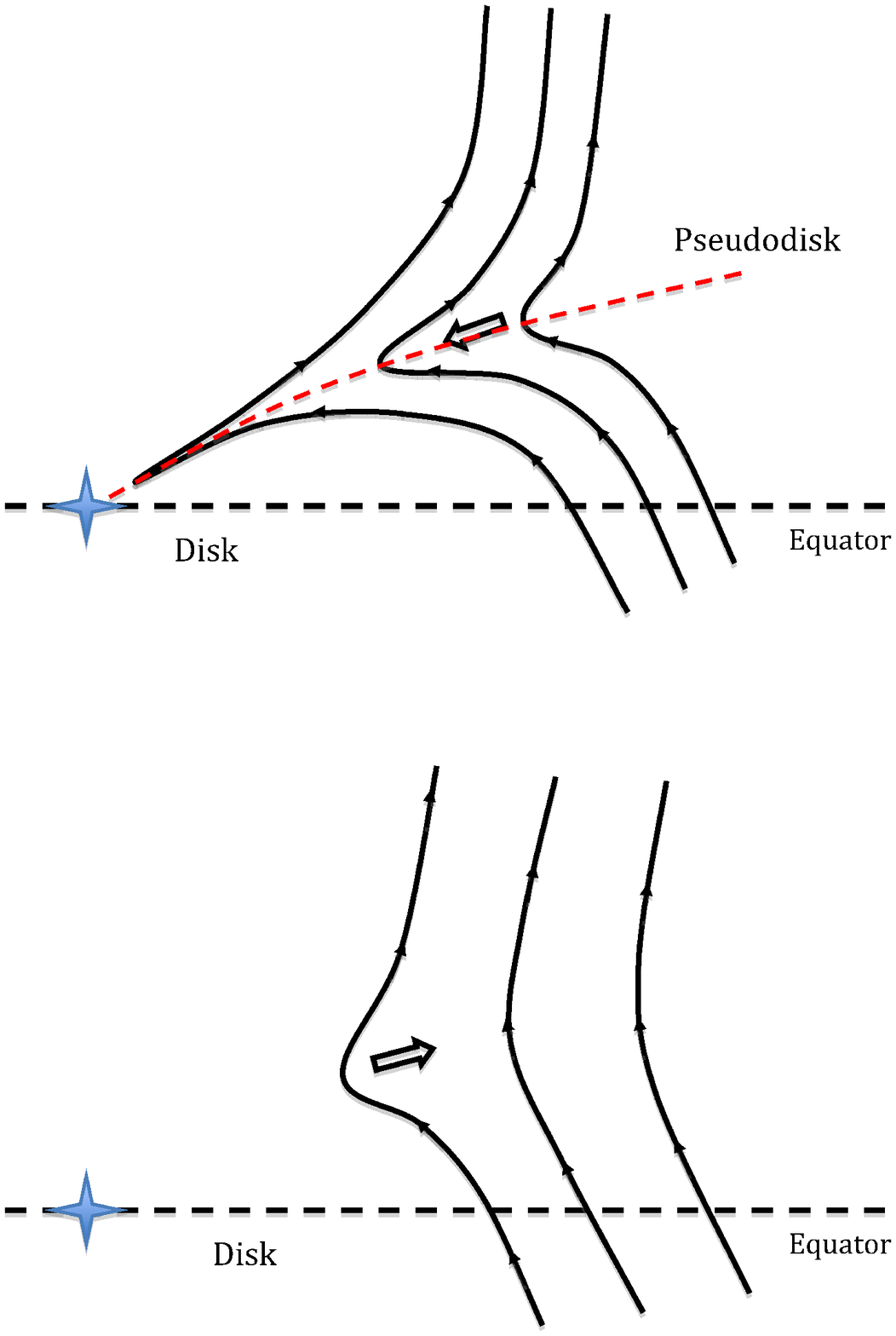}
\caption{Cartoon illustrating pseudodisk formation and magnetic flux
loss from the equatorial, disk-forming region close to the central
object. Top panel: localized runaway gravitational collapse drags
the field lines into a highly pinched configuration, which enables
matter to slide along field lines and collect at the apex to form a
dense pseudodisk, which is further compressed magnetically.
The out-of-the-equatorial-plane warping enables the pseudodisk
to deliver mass close to the center object without increasing the
magnetic flux in the circumstellar disk-forming region on the
equatorial plane. Bottom panel: the highly pinched field lines
reconnect near the central object, triggered by magnetic decoupling
near the central object or some other means.
The reconnected field lines are driven outward by the
magnetic tension, escaping to large distances.
}
\label{cartoon}
\end{figure}

The pseudodisk that plays a central role in our scenario of RSD formation
is a generic feature of the protostellar collapse channeled by a
large-scale, dynamically significant magnetic field. This is
because the material distributed along any given field line
cannot all collapse toward the center at the same rate; some
part is bound to collapse in a runaway fashion, as illustrated
in the upper panel of Fig.\ \ref{cartoon}. If a piece of matter
on a field line is initially closer to the central object
than the rest of the material along the same field line, it
would experience a stronger gravitational acceleration, which
would move it closer to the center, which would in turn increase
its gravitational acceleration further. This differential
collapse of matter along a field line drags the field line
into a highly pinched configuration, which would not only
allow matter to slide along the field
line to the cusp but also compress the material collected there
into a flattened structure --- the pseudodisk (\ct{GalliShu1993};
\ct{Allen+2003}). It is the conduit for most of the core
mass accretion with or without
a turbulence,\footnote{In the most general case, the
gravity-driven, magnetically channeled dense thin accretion
regions may appear as a network of dense, collapsing ribbons rather than a
single topologically connected structure. They are a generalized form
of the pseudodisk.} as discussed in \S\ \ref{Corrugation} and
illustrated in Fig.\ \ref{DenSphere}. As such, it is
largely responsible for concentrating the magnetic flux at small
radii that creates the difficulty for RSD formation in the
first place. Fortunately, it also holds the key to overcoming the difficulty.

The key is the sharp field reversal across the pseudodisk (see
Fig.\ \ref{cartoon}). The highly pinched field lines are prone
to magnetic reconnection, numerical or otherwise. There is
little doubt that reconnection has occurred in all magnetized disk
formation simulations to date that include turbulence
(\ct{Santos-Lima+2012,Santos-Lima+2013}; \ct{Seifried+2012,Seifried+2013};
\ct{Joos+2013}; \ct{Myers+2013}). It is needed to explain the
loss of magnetic flux near the central protostar relative to that
expected under flux-freezing found in these simulations.
\citet{Santos-Lima+2012} was the first to study the flux loss, and
attributed it to the turbulence-induced magnetic reconnection
(\ct{LazarianVishniac1999}; \ct{Kowal+2009}), which in
their scenario is the key to disk formation (see also
\ct{Santos-Lima+2013}). \citet{Joos+2013} found that the amount of
flux loss increases with the level of turbulence, as one would
expect if the flux loss is induced by turbulent reconnection. This
is, however, not definitive proof of \citeauthor{Santos-Lima+2013}'s
scenario, because the turbulence-induced reconnection events
remain to be identified in the simulations.

We propose an alternative scenario for the reconnection that is
required to explain the flux reduction observed in simulations,
including our own: the magnetic decoupling-triggered
reconnection of sharply pinched field lines. This alternative is
motivated by the fact that gravitational collapse can naturally
produce, by itself, sharply pinched field lines close to the central
object that are prone to reconnection and that flux reduction is
observed even in non-turbulent simulations, such as our Models A and P (see
also \ct{Zhao+2011} and \ct{Krasnopolsky+2012}), which indicates
that efficient reconnection can be achieved
without any turbulence. This type of reconnection was observed
directly in 2D (axisymmetric) simulations of \citeauthor{MellonLi2008}
(\citeyear{MellonLi2008},
see also footnote \ref{2D}, Fig.\ \ref{2DSnap}, and auxiliary material online), where
oppositely directed field lines above
and below the pseudodisk reconnect episodically near the inner
boundary, where the matter is decoupled from the field lines as
it accretes onto the central object. As discussed in
\S\ \ref{MagFlux}, the decoupling is required for solving the
``magnetic flux problem'' in star formation, and may be achieved
physically through non-ideal MHD effects (e.g., \ct{LiMcKee1996};
\ct{Contopoulos+1998}; \ct{KunzMouschovias2010}; \ct{Machida+2011};
\ct{DappBasu2010}; \ct{Dapp+2012}; \ct{Tomida+2013}).
In our scenario, it is responsible
for preventing the field lines from piling up near the center to
form the strong split magnetic monopole that lies at the heart of
the ``magnetic braking catastrophe'' in
ideal MHD (\ct{Galli+2006}).\footnote{We note that, once a RSD
has formed, its differential rotation can force field lines of
opposite polarity closer and closer together, which can also trigger
reconnection, as noted in \citeauthor{Li+2013} (\citeyear{Li+2013}; see the left panel of
their Fig.\ 6). The rotation-induced reconnection may help the RSD
survive by decreasing the level of its magnetization.
}

The elimination of the split monopole does not guarantee RSD
formation, however.
In the absence of any turbulence (or field-rotation misalignment),
the bulk of core mass accretion is funneled through a dense,
coherent, equatorial pseudodisk (see Fig.\ \ref{Pseudo}).
The accreting material drags the field lines to the inner boundary,
where they decouple from the matter. After decoupling, the
oppositely directing field lines above and below the equator
reconnect, and are driven outward by the magnetic tension force
along the equatorial plane. However, their equatorial escape to
large distances is blocked by continuous mass infall in the dense,
coherent, equatorial pseudodisk. They remain trapped close to the
central object, in a highly magnetized circumstellar region ---
the DEMS\@. As discussed in \S\ \ref{dems_discussion} and illustrated
in Fig.\ \ref{DEMS}, the DEMS must be removed in order for
a robust, rotationally supported disk to form (\ct{Zhao+2011};
\ct{Krasnopolsky+2012}). This, in our scenario, is where
turbulence (and field-rotation misalignment) comes in.

In the presence of a strong turbulence, the pseudodisk can become
severely warped out of the equatorial plane and highly variable
in time. The beneficial effect of pseudodisk warping to disk
formation is illustrated in the top panel of the cartoon in
Fig.\ \ref{cartoon}. Mass accretion through the warped pseudodisk
will still drag along the (highly pinched) field lines. However,
unlike the case of flat equatorial pseudodisk, such field lines
do not have to pass through the circumstellar, disk-forming region
on the equatorial plane; they can cross the equatorial plane at
larger distances. The situation is qualitatively similar in the
presence of a large field-rotation misalignment, which
warps the plane of pseudodisk away from the plane of disk
formation. In both cases, when the highly pinched field lines
threading the warped pseudodisk reconnect, they can escape directly
to large distances without having to cross the equatorial
disk-forming region first (see the lower panel of
Fig.\ \ref{cartoon}). As a result, the amount of magnetic flux
trapped in the equatorial, disk-forming region is much reduced
compared to the non-turbulent, field-rotation aligned case.
The reduction greatly weakens the DEMS, making the disk formation
possible.

Our proposed scenario of RSD formation in turbulent magnetized dense
cores thus involves two conceptually distinct steps:
(1) decoupling-triggered reconnection of sharply pinched field lines close to the protostar, which
removes the strong split magnetic monopole at the center, the first
obstacle to disk formation, and (2) warping of the pseudodisk out of
the disk-forming plane, which weakens the DEMS, the second obstacle
to disk formation. Compared to Santos-Lima et al.'s scenario of
turbulence-induced reconnection, it has the advantage of being capable
of explaining the disk formation enabled by both turbulence and
field-rotation misalignment. Nevertheless, the two scenarios are not mutually
exclusive. Indeed, it is likely that both mechanisms are operating in
the current generation of simulations. For example, field-matter
decoupling must be present in any magnetized disk formation
simulations involving sink particles, including those of
\citet{Santos-Lima+2012,Santos-Lima+2013},
because the matter is accreted onto the sink particle
but not the magnetic field. On the other hand, turbulence has been shown
to enhance the reconnection rate of oppositely directed field lines,
both analytically (\ct{LazarianVishniac1999}) and numerically
(e.g., \ct{Kowal+2009}), so the turbulence-induced reconnection
is likely present in simulations, including our own, although its
rate is difficult to quantify. We should stress that, even in
Santos-Lima et al.'s scenario, the pseudodisk is expected to
play a central role: its
sharply pinched field lines make it the most natural location for
the turbulence-induced reconnection. Furthermore, the warping of
the pseudodisk, a key ingredient of our scenario, can help
such reconnected field lines escape to large distances without
passing through (and being trapped in) the equatorial, disk-forming
region. One complication
is that the turbulent motions are expected to be strongly modified,
indeed dominated, by supersonic gravitational infall in the
pseudodisk region close to the central object. The potential
effect of such fast infall on the turbulence-induced magnetic
reconnection remains to be quantified.
Another complication is that, in ideal MHD simulations, both
turbulence-induced and decoupling-triggered reconnections involve
numerical diffusion, which depends on numerical resolution. As such,
it would be difficult to obtain numerically converged solutions.

\subsection{Characteristics of Disks Fed by Warped, Magnetized Pseudodisks}
\label{DiskCharac}

\begin{figure}
\epsscale{1.2}
\plottwo{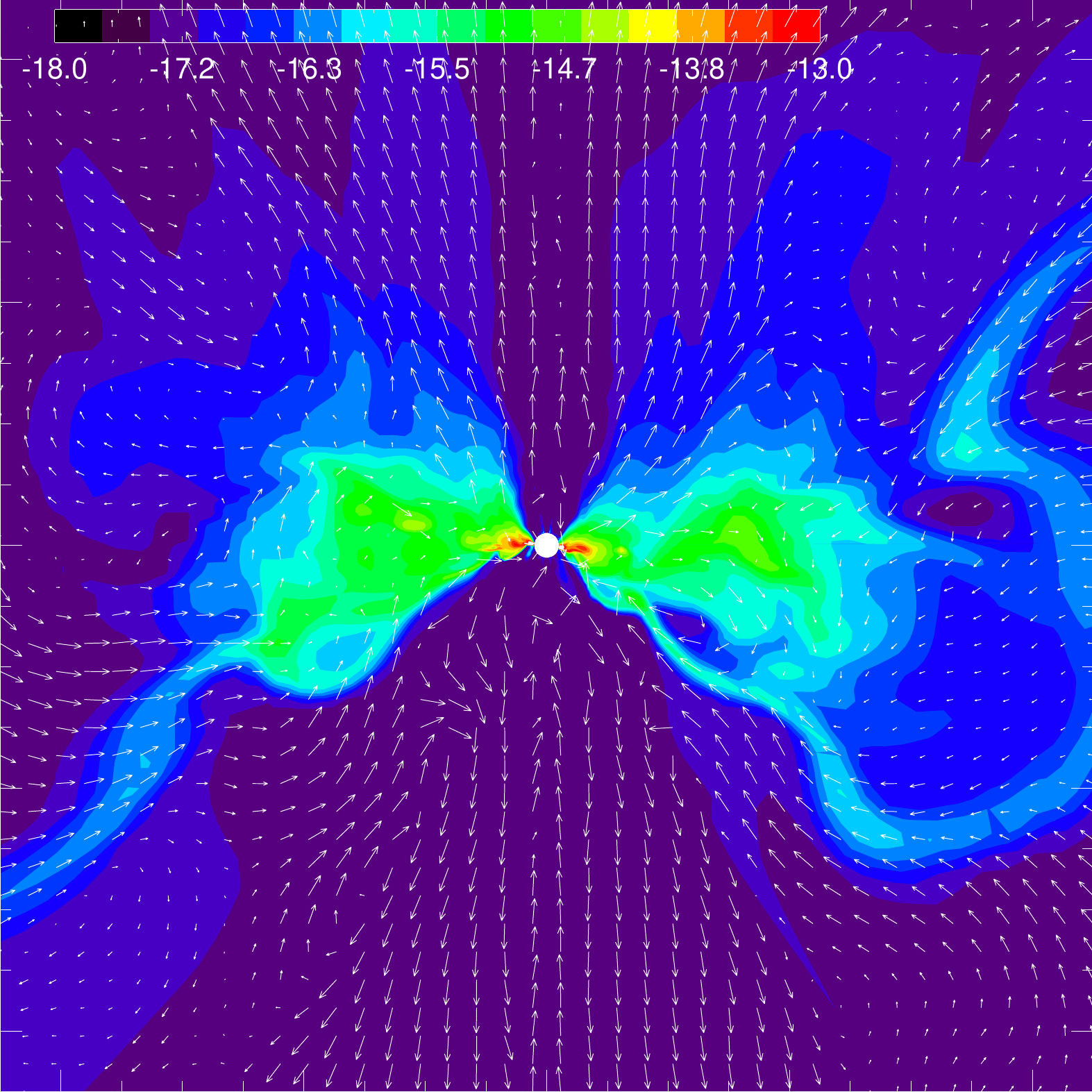}{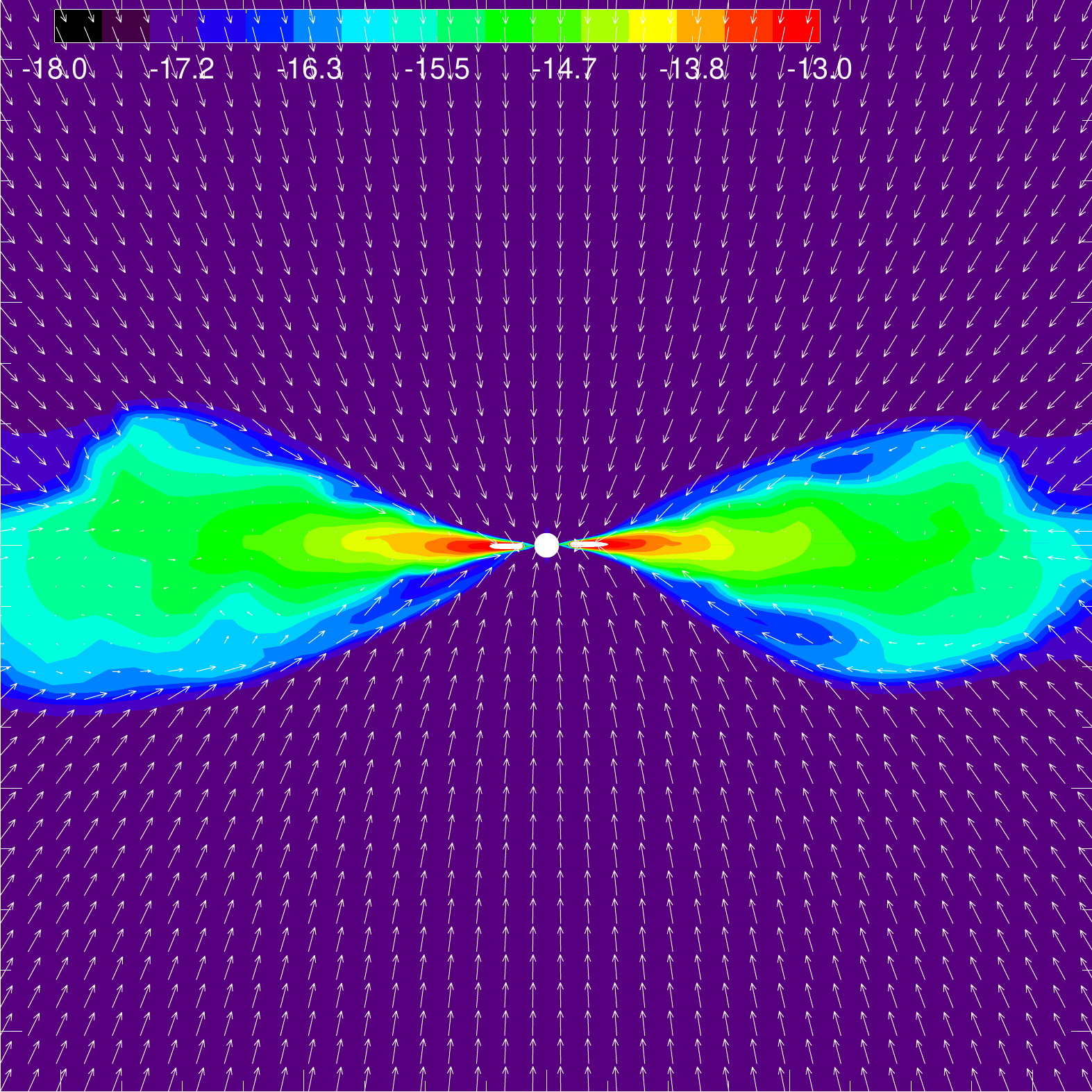}
\caption{Disk comparison. Plotted are the distribution of logarithm of
density and velocity field on a representative meridian plane for
the magnetized, turbulent case of Model F ($M=1$, left panel) and
its non-magnetic and non-turbulent counterpart (Model H, right
panel) at the same time ($t=8\times 10^{11}\second$). Note that the
disk in the former appears thicker and more dynamically active than in
the latter. The length of the box is $9\times 10^{15}\cm$ (or
$600\au$) on each side. The colorbar is as in Fig.\ \ref{TurbE_6p}.
An axial dense spot has been erased for better clarity of the equatorial region.
The velocity vectors are plotted with the
maximum speed capped at $10^5\cms$, also for clarity.
}
\label{Disk}
\end{figure}

A key finding of our investigation is that the rotationally
supported disks formed in turbulent, magnetized cloud cores
are fed by highly variable, strongly warped pseudodisks.
An interesting characteristic of such disks is their thickness.
It is illustrated in
the left panel of Fig.\ \ref{Disk} for Model F (with a turbulent
Mach number $M=1$); the figure is a zoom-in of Fig.\ \ref{WarpB}
(see also the lower panels of Fig.\ \ref{power} for Models U and V)\@.
For comparison, we also plotted side-by-side the disk formed in
a model that is neither magnetic nor turbulent, but with other
parameters identical to those of Model F (Model H in Table 1).
The disk in the magnetized Model F is smaller in radius and thicker
(relative to radius) than that in the hydro Model H\@. The smaller
radius is to be expected because of angular momentum removal
by magnetic braking and the associated outflow. The larger
thickness may be due, at least in part, to the feeding of the disk
by a strongly warped pseudodisk from directions that are
highly variable and often tilted significantly away from the
equatorial plane (see auxiliary material online for a movie of mass
accretion in a meridian plane); indeed, the warped
pseudodisk often feeds the rotationally support disk from the top and
bottom surfaces, rather than the outer edge of the disk. As a result,
the disk is dynamically ``hotter'' (with faster motions) in the poloidal
plane than that in the hydro case (compare the disk velocity
fields in Fig.\ \ref{Disk}; see below for another mechanism for
puffing up the disk).

\begin{figure}
\epsscale{1.0}
\plotone{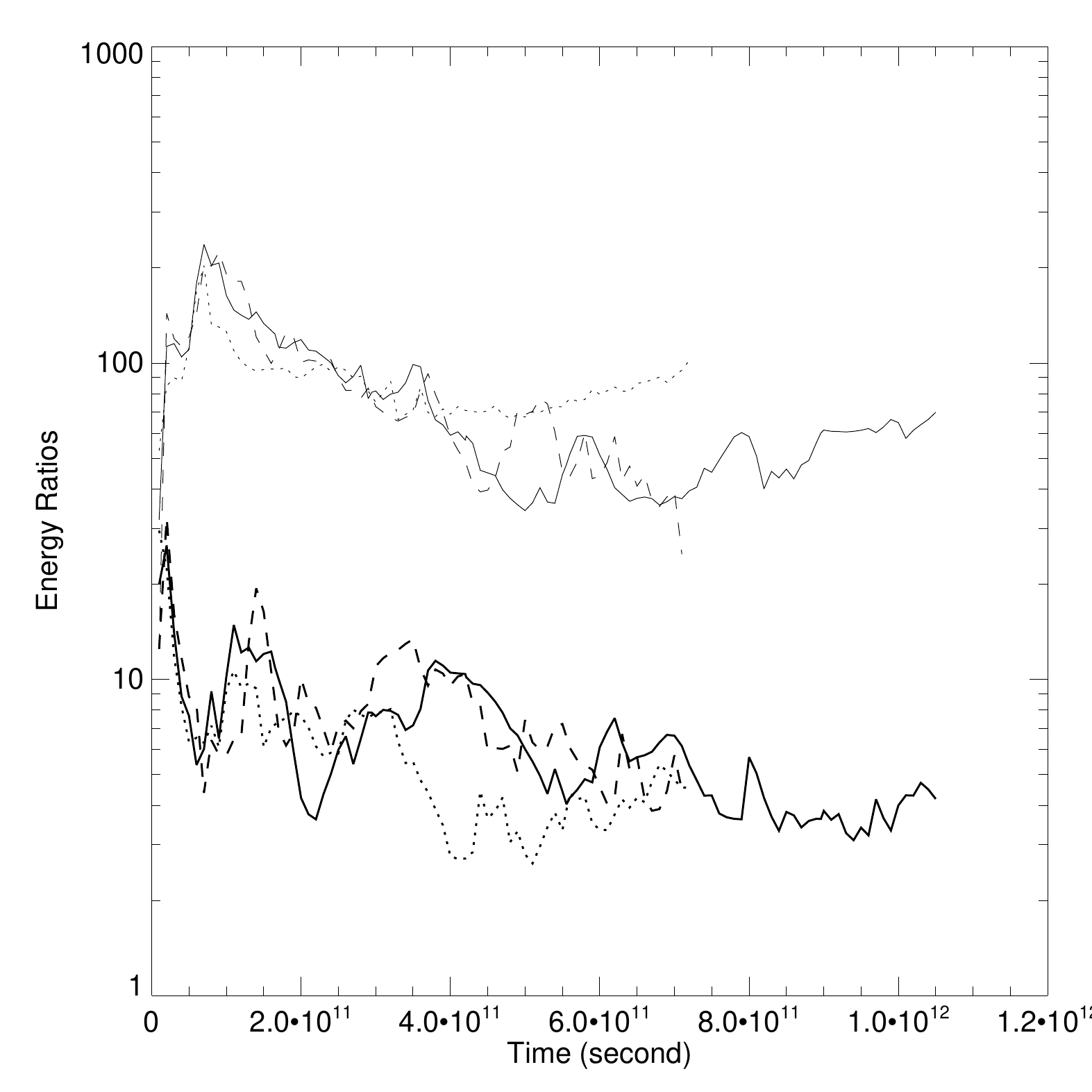}
\caption{Time evolution of the ratios of the magnetic to thermal
(lower thicker lines) and rotational to thermal (upper thinner
lines) energies for the
disk in Model F ($p=1$, solid lines), U ($p=0.5$, dashed) and V
($p=2.0$, dotted), showing that the disk remains strongly magnetized.
}
\label{Ratio}
\end{figure}

\begin{figure}
\epsscale{0.8}
\plotone{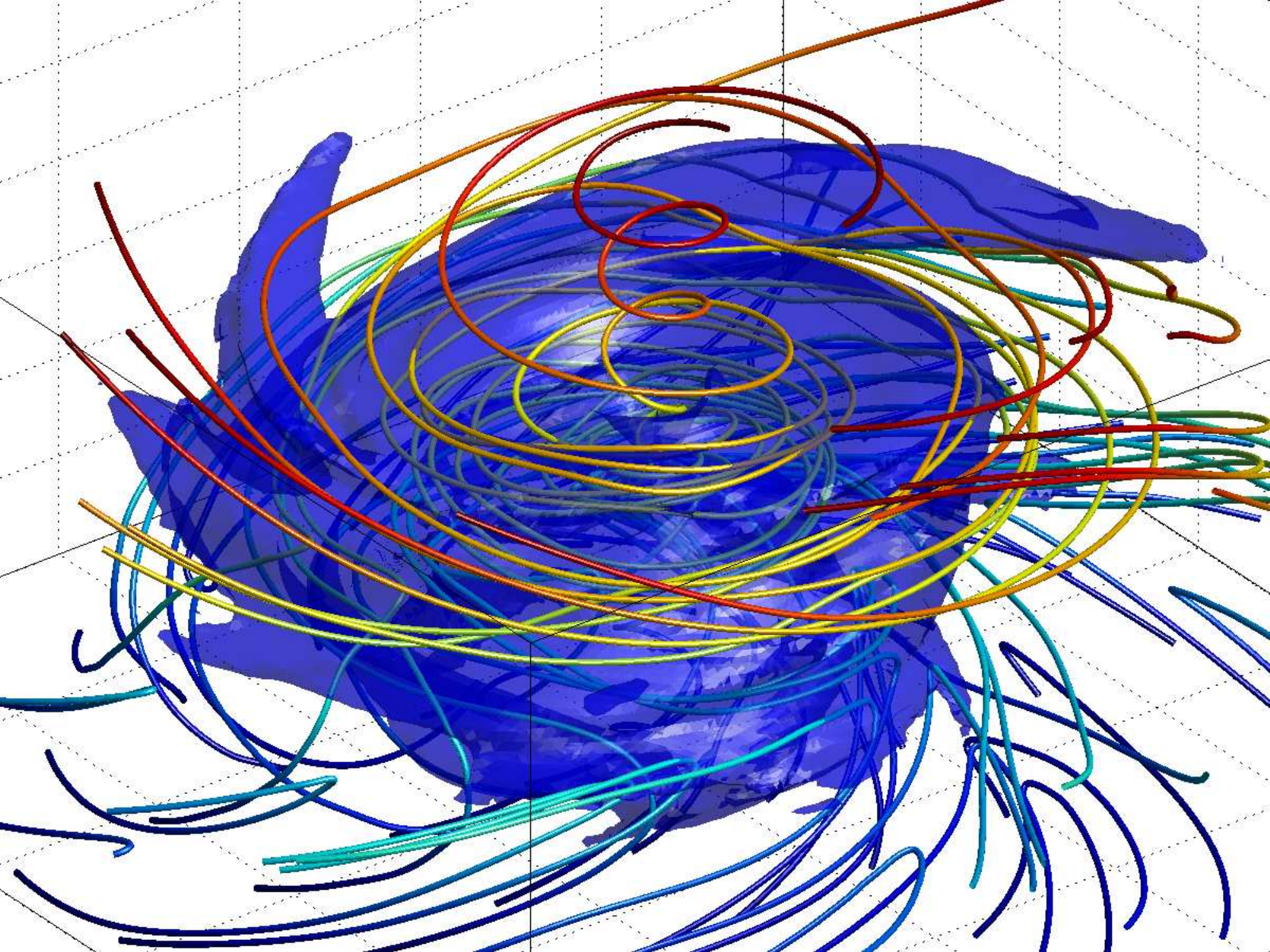}
\caption{3D magnetic field structure of the disk shown in the left
panel of Fig.\ \ref{Disk}. Plotted are representative magnetic field
lines and isodensity surface at $10^{-16}\rhounit$ (blue). The
box size is $600 \times 300\au$.
}
\label{DiskB}
\end{figure}

Another characteristic of the disks fed by warped pseudodisks is
that they are significantly magnetized. This is because the
pseudodisks are necessarily magnetized to a significant level
since they are the product of magnetically-channeled gravitational
collapse. Significant magnetization is therefore expected for the
disks as well. The degree of disk magnetization is shown in
Fig.\ \ref{Ratio}, where we plot the time evolution of the ratios
of magnetic to thermal and rotational to thermal energies for
the disk (defined somewhat crudely as the region denser than
$10^{-16}\rhounit$, a density corresponding to the blue
isodensity surface in Fig.\ \ref{3DView}) for the sonic turbulence
cases (Models F, U and V)\@. It is clear that the
magnetic energy dominates the thermal energy, by a factor of a few
to several for all three cases. The magnetic energy is less than
the rotational energy,
however, by about one order of magnitude. The magnetic field is
therefore expected to be wrapped by rotation into a predominantly
toroidal configuration. This is indeed the case, as illustrated
in Fig.\ \ref{DiskB}, where representative magnetic field lines
are plotted for the disk of Model F shown in the left panel of
Fig.\ \ref{Disk}.
The toroidal field configuration is consistent with the recent
dust polarization observations of the young disk in IRAS 16293B
(\ct{Rao+2014}), HL Tau (\ct{Stephens+2014}) and L1527
(\ct{Segura-Cox+2014}).

The two characteristics of the disks discussed above (large thickness
and significant magnetization) may be related. The rather strong
toroidal field inside the disk provides an additional support (on
top of the thermal pressure) to the gas in the vertical direction,
which tends to puff up the
disk. Observational evidence for a puffed-up disk may already exist.
\citet{Tobin+2013} inferred, through detailed modeling of the
L$^\prime$-band image, that the young disk in L1527 is thicker than those
in more evolved sources on the $100\au$ scale; it is about twice
the disk scale-height estimated based on the thermal support alone.
To puff up a disk by a factor of 2, an additional (non-thermal)
energy density of $\sim 3$ times the thermal energy density is
needed, since the scale-height is proportional to the square root
of the total (thermal and non-thermal) energy density. From
Fig.\ \ref{Ratio},
it is clear that the required non-thermal energy is comparable to
the magnetic energy in the disk. It is therefore plausible that
the puffed-up disk in L1527 is an example of the kind of thick,
dynamically active disks fed anisotropically by highly variable,
strongly warped, magnetized pseudodisks that we find in our
simulations (and possibly in
the simulations of \ct{Santos-Lima+2012,Santos-Lima+2013};
\ct{Seifried+2012,Seifried+2013};
\ct{Myers+2013} and \ct{Joos+2013} as well).
High resolution observations of polarized dust emission from
the disks of L1527 and other deeply embedded sources using
sub/millimeter interferometers (especially ALMA) can help
firm up or refute this interpretation.
In any case, the disk thickness compared to the
thermal scale-height can put an upper limit on the disk
toroidal field strength, which is difficult to
constrain through other means.

\subsection{Implications, Uncertainties and Future Directions}
\label{caveat}

The picture of disk-feeding by a variable, warped magnetized
pseudodisk, if true in general, may
have strong implications for the chemical connection between
the collapsing core and the disk; the connection is an important
step toward understanding the chemical heritage of the solar
system (e.g., \ct{CaselliCeccarelli2012}; \ct{Hincelin+2013}).
First, if most of the disk-forming material comes from the
magnetically compressed pseudodisk, its density before
entering the RSD should be higher than that in the non-magnetic
(hydro) case. The higher density
could affect the rates of chemical reactions and ice formation (for
example, through shorter adsorption timescales, or perhaps three-body
reactions if the density is high enough),
and thus the gas and ice composition. Second, disk-feeding
through a highly variable pseudodisk means that any accretion shock,
if exists at all, is strongly time dependent and spatially localized,
unlike the simplest hydro case where a well-defined accretion shock
encases the whole disk (see, e.g., \ct{Yorke+1993}, and the right
panel of Fig.\ \ref{Disk}).
The shock structure is expected to
be further modified by the magnetic field embedded in the
pseudodisk. As a result, the disk material may experience
a rather different thermal history, which could affect both
its gas and ice content (e.g., \ct{Visser+2009}).
Third, if the disk is puffed up and dynamically active in the poloidal
plane (see the left panel of Fig.\ \ref{Disk}), its rates of chemical
reactions and vertical mixing would be affected.
Furthermore, the long-term evolution of the disk is expected to be
strongly modified, perhaps dominated, by the rather strong (toroidal)
magnetic field in the disk. A caveat is that the disk magnetic energy may be
strongly affected by non-ideal MHD effects, which are expected to
be important since the bulk of protostellar disks is lightly ionized
(\ct{Armitage11}; \ct{Turner+2014}). The modifications need
to be quantified in the future.

Another caveat is that, in our simulations, we included the gravity
from a central object (of $0.5\msun$) but not the self-gravity
of the gas. One consequence of this idealization is that the gravity
is stronger at small radii compared to the more self-consistent case
with self-gravity before the central object accretes $0.5\msun$.
The stronger gravity is expected to accelerate the material in the
pseudodisk to a higher infall speed relative to the more slowly
collapsing material at larger distances that is magnetically
connected to it. The higher relative speed is expected to stretch
the field lines across the pseudodisk into a more severely pinched
configuration, which should in turn compress the pseudodisk to a
smaller thickness. This has the benefit of bringing the role of
pseudodisk on disk formation into a sharper focus, but it may have
exaggerated that role somewhat. Nevertheless, the presence of a
pseudodisk in self-gravitating magnetized protostellar collapse is well
established. The fact that we are able to reproduce the known
results that disk formation in non-turbulent cores is suppressed
when the magnetic field and rotation axis are aligned (Model A)
and enabled when they are orthogonal (Model P) gives us confidence
that, despite the idealized setup, our results are qualitatively
correct. It remains to be determined whether our quantitative results,
such as the RSD formation enabled by a sonic ($M=1$) turbulence in a
$\lambda=2.92$ core (e.g., Model F), hold up or not when the
self-gravity is included.

We should note that the turbulence adopted in our simulations is
somewhat ad hoc. It serves well the purposes of perturbing the
pseudodisk and enabling disk formation, but how closely it
resembles the real turbulence in dense cores of molecular
clouds is unclear. This drawback will be harder to remedy,
because the detailed properties of the turbulence, such as
its energy spectrum, are not well quantified observationally
on the core scale, although the situation should improve with
high-resolution ALMA observations.

\section{Summary}
\label{conclusion}

We have carried out idealized numerical experiments of the accretion of a
rotating, turbulent, but non-self-gravitating, dense core onto a
pre-existing central stellar object in the presence of a moderately
strong magnetic field.
We found that, in agreement with previous work, the formation of a
rotationally supported disk (RSD) is suppressed by the magnetic
field in the absence of any turbulence (or field-rotation
misalignment) and that an initial turbulence, if strong enough,
can enable RSD formation. We identified the physically motivated
magnetic decoupling-triggered reconnection of severely pinched field
lines close to the central object and the warping of
the pseudodisk out of the disk-forming, equatorial plane as two
key ingredients of the turbulence-enabled disk formation, in
contrast to the previously suggested scenario that relies exclusively
on turbulence-induced reconnection; in our picture, the field pinching
that facilitates the reconnection arises primarily from (anisotropic)
gravitational collapse rather than turbulence (see Fig.\ \ref{cartoon}).
The decoupling-triggered reconnection weakens the split magnetic
monopole near the protostar, which is the first obstacle to disk
formation in a magnetized cloud core. The turbulence-induced
pseudodisk warping weakens the so-called ``magnetic decoupling
enabled structure'' (DEMS), the second obstacle to disk formation,
by reducing the amount of the magnetic flux trapped in the equatorial,
disk-forming region. We also showed that the warping
decreases the rates of angular momentum removal from the infalling
material in the pseudodisk region by both magnetic torque (especially
near the so-called ``magnetic barrier,'' see Fig.\ \ref{Pseudo}) and
outflow, leaving more angular momentum to form a rotationally
supported disk. The beneficial effects of warping the pseudodisk
out of the disk-forming (equatorial) plane can also be achieved
by a misalignment between the magnetic field and rotation axis. In
this sense, the turbulence- and misalignment-enabled disk formation
mechanisms are unified.

We emphasized that the pseudodisk is an unavoidable product of the
highly anisotropic, magnetically-channeled gravitational collapse,
even in the presence of turbulence. It is the main conduit for core
mass accretion and its severely pinched field configuration makes
it a natural place for the magnetic reconnection triggered by
decoupling of both physical and numerical origin and possibly enhanced
by turbulence. It feeds the rotationally
supported disks formed in our turbulent, magnetized dense cores.
These disks differ significantly from those formed in the
non-turbulent, non-magnetic cores. They are thicker, more
dynamically active in the poloidal plane, fairly
strongly magnetized, and are not completely encased by an
accretion shock.
It will be interesting to determine whether
these differences persist when the self-gravity of the material
surrounding the stellar object is included and, if yes, to
explore their implications, especially on the disk chemistry
(including ice) and their long-term dynamical evolution
(including possible fragmentation and substellar object formation).

\acknowledgements
We thank Ugo Hincelin for useful discussion. The work is supported in part
by NNX10AH30G, NNX14AB38G, and AST1313083.

\end{document}